\documentclass[prc,10pt,twocolumn,tightenlines,superscriptaddress]{revtex4-1}
\usepackage{natbib}     %bibliography

\usepackage{hyperref}   % use for hypertext links, including those to external documents and URLs
\usepackage{graphicx}   % need for figures
\usepackage{verbatim}   % useful for program listings
\usepackage{color}      % use if color is used in text
\usepackage{placeins}   % for figure control
\usepackage{amsmath}    % for splitting equations across multiple lines
\graphicspath{{Figures/}}
\usepackage{units}   %for \nicefrac?

\usepackage{array,longtable}%longtable, can flow into multiple pages
\AtBeginDocument{\usepackage{booktabs}}   

\begin{document}

\title{Measurement of the Differential and Total Cross Sections of the $\gamma d \to K^{0}\Lambda (p)$ Reaction within the Resonance Region}

%%%%%%%%%%%%%%% Latex Macros for institute addresses  %%%%%%%%%%%%%%%%%%%%%%%%% 

\newcommand*{\ANL}{Argonne National Laboratory, Argonne, Illinois 60439}
\newcommand*{\ANLindex}{1}
\affiliation{\ANL}
\newcommand*{\ASU}{Arizona State University, Tempe, Arizona 85287-1504}
\newcommand*{\ASUindex}{2}
\affiliation{\ASU}
\newcommand*{\BONN}{Helmholtz-Institute f\"ur Strahlen- und Kernphysik der Rheinischen Friedrich-Wilhelms Universit\"at Bonn}
\newcommand*{\BONNindex}{3}
\affiliation{\BONN}
\newcommand*{\CSUDH}{California State University, Dominguez Hills, Carson, CA 90747}
\newcommand*{\CSUDHindex}{4}
\affiliation{\CSUDH}
\newcommand*{\CANISIUS}{Canisius College, Buffalo, NY}
\newcommand*{\CANISIUSindex}{5}
\affiliation{\CANISIUS}
\newcommand*{\CMU}{Carnegie Mellon University, Pittsburgh, Pennsylvania 15213}
\newcommand*{\CMUindex}{6}
\affiliation{\CMU}
\newcommand*{\CUA}{Catholic University of America, Washington, D.C. 20064}
\newcommand*{\CUAindex}{7}
\affiliation{\CUA}
\newcommand*{\SACLAY}{Irfu/SPhN, CEA, Universit\'e Paris-Saclay, 91191 Gif-sur-Yvette, France}
\newcommand*{\SACLAYindex}{8}
\affiliation{\SACLAY}
\newcommand*{\CNU}{Christopher Newport University, Newport News, Virginia 23606}
\newcommand*{\CNUindex}{9}
\affiliation{\CNU}
\newcommand*{\UCONN}{University of Connecticut, Storrs, Connecticut 06269}
\newcommand*{\UCONNindex}{10}
\affiliation{\UCONN}
\newcommand*{\FU}{Fairfield University, Fairfield CT 06824}
\newcommand*{\FUindex}{11}
\affiliation{\FU}
\newcommand*{\FERRARAU}{Universita' di Ferrara , 44121 Ferrara, Italy}
\newcommand*{\FERRARAUindex}{12}
\affiliation{\FERRARAU}
\newcommand*{\FIU}{Florida International University, Miami, Florida 33199}
\newcommand*{\FIUindex}{13}
\affiliation{\FIU}
\newcommand*{\FSU}{Florida State University, Tallahassee, Florida 32306}
\newcommand*{\FSUindex}{14}
\affiliation{\FSU}
\newcommand*{\Genova}{Universit$\grave{a}$ di Genova, 16146 Genova, Italy}
\newcommand*{\Genovaindex}{15}
\affiliation{\Genova}
\newcommand*{\GWUI}{The George Washington University, Washington, DC 20052}
\newcommand*{\GWUIindex}{16}
\affiliation{\GWUI}
\newcommand*{\ISU}{Idaho State University, Pocatello, Idaho 83209}
\newcommand*{\ISUindex}{17}
\affiliation{\ISU}
\newcommand*{\INFNFE}{INFN, Sezione di Ferrara, 44100 Ferrara, Italy}
\newcommand*{\INFNFEindex}{18}
\affiliation{\INFNFE}
\newcommand*{\INFNFR}{INFN, Laboratori Nazionali di Frascati, 00044 Frascati, Italy}
\newcommand*{\INFNFRindex}{19}
\affiliation{\INFNFR}
\newcommand*{\INFNGE}{INFN, Sezione di Genova, 16146 Genova, Italy}
\newcommand*{\INFNGEindex}{20}
\affiliation{\INFNGE}
\newcommand*{\INFNRO}{INFN, Sezione di Roma Tor Vergata, 00133 Rome, Italy}
\newcommand*{\INFNROindex}{21}
\affiliation{\INFNRO}
\newcommand*{\INFNTUR}{INFN, Sezione di Torino, 10125 Torino, Italy}
\newcommand*{\INFNTURindex}{22}
\affiliation{\INFNTUR}
\newcommand*{\ORSAY}{Institut de Physique Nucl\'eaire, CNRS/IN2P3 and Universit\'e Paris Sud, Orsay, France}
\newcommand*{\ORSAYindex}{23}
\affiliation{\ORSAY}
\newcommand*{\ITEP}{Institute of Theoretical and Experimental Physics, Moscow, 117259, Russia}
\newcommand*{\ITEPindex}{24}
\affiliation{\ITEP}
\newcommand*{\JMU}{James Madison University, Harrisonburg, Virginia 22807}
\newcommand*{\JMUindex}{25}
\affiliation{\JMU}
\newcommand*{\KNU}{Kyungpook National University, Daegu 41566, Republic of Korea}
\newcommand*{\KNUindex}{26}
\affiliation{\KNU}
\newcommand*{\MISS}{Mississippi State University, Mississippi State, MS 39762-5167}
\newcommand*{\MISSindex}{27}
\affiliation{\MISS}
\newcommand*{\UNH}{University of New Hampshire, Durham, New Hampshire 03824-3568}
\newcommand*{\UNHindex}{28}
\affiliation{\UNH}
\newcommand*{\NSU}{Norfolk State University, Norfolk, Virginia 23504}
\newcommand*{\NSUindex}{29}
\affiliation{\NSU}
\newcommand*{\OHIOU}{Ohio University, Athens, Ohio  45701}
\newcommand*{\OHIOUindex}{30}
\affiliation{\OHIOU}
\newcommand*{\ODU}{Old Dominion University, Norfolk, Virginia 23529}
\newcommand*{\ODUindex}{31}
\affiliation{\ODU}
\newcommand*{\PNPI}{Petersburg Nuclear Physics Institute}
\newcommand*{\PNPIindex}{32}
\affiliation{\PNPI}
\newcommand*{\RPI}{Rensselaer Polytechnic Institute, Troy, New York 12180-3590}
\newcommand*{\RPIindex}{33}
\affiliation{\RPI}
\newcommand*{\ROMAII}{Universita' di Roma Tor Vergata, 00133 Rome Italy}
\newcommand*{\ROMAIIindex}{34}
\affiliation{\ROMAII}
\newcommand*{\MSU}{Skobeltsyn Institute of Nuclear Physics, Lomonosov Moscow State University, 119234 Moscow, Russia}
\newcommand*{\MSUindex}{35}
\affiliation{\MSU}
\newcommand*{\SCAROLINA}{University of South Carolina, Columbia, South Carolina 29208}
\newcommand*{\SCAROLINAindex}{36}
\affiliation{\SCAROLINA}
\newcommand*{\TEMPLE}{Temple University,  Philadelphia, PA 19122 }
\newcommand*{\TEMPLEindex}{37}
\affiliation{\TEMPLE}
\newcommand*{\JLAB}{Thomas Jefferson National Accelerator Facility, Newport News, Virginia 23606}
\newcommand*{\JLABindex}{38}
\affiliation{\JLAB}
\newcommand*{\UTFSM}{Universidad T\'{e}cnica Federico Santa Mar\'{i}a, Casilla 110-V Valpara\'{i}so, Chile}
\newcommand*{\UTFSMindex}{39}
\affiliation{\UTFSM}
\newcommand*{\EDINBURGH}{Edinburgh University, Edinburgh EH9 3JZ, United Kingdom}
\newcommand*{\EDINBURGHindex}{40}
\affiliation{\EDINBURGH}
\newcommand*{\GLASGOW}{University of Glasgow, Glasgow G12 8QQ, United Kingdom}
\newcommand*{\GLASGOWindex}{41}
\affiliation{\GLASGOW}
\newcommand*{\VT}{Virginia Tech, Blacksburg, Virginia   24061-0435}
\newcommand*{\VTindex}{42}
\affiliation{\VT}
\newcommand*{\VIRGINIA}{University of Virginia, Charlottesville, Virginia 22901}
\newcommand*{\VIRGINIAindex}{43}
\affiliation{\VIRGINIA}
\newcommand*{\WM}{College of William and Mary, Williamsburg, Virginia 23187-8795}
\newcommand*{\WMindex}{44}
\affiliation{\WM}
\newcommand*{\YEREVAN}{Yerevan Physics Institute, 375036 Yerevan, Armenia}
\newcommand*{\YEREVANindex}{45}
\affiliation{\YEREVAN}
 
\newcommand*{\NOWHAMPTON}{Hampton University, Hampton, VA 23668}
\newcommand*{\NOWDAMMAM}{University of Dammam College of Education of Jubail Department of Physics P.O 12020, Industrial Jubail 31961 Saudi Arabia}

 %%%%%%%%%%%%%%% END OF Latex Macros for institute addresses  %%%%%%%%%%%%%%%%%%%%%%%%% 

\author{N. Compton}
\affiliation{\OHIOU}
\author{C.E. Taylor}
\affiliation{\ISU}
\author{K. Hicks}
\affiliation{\OHIOU}
\author {P. Cole}
\affiliation{\ISU}
\author{N. Zachariou}
\affiliation{\EDINBURGH}
\author{Y. Ilieva}
\affiliation{\SCAROLINA}
\author{P. Nadel-Turonski}
\affiliation{\JLAB}

\author{E. Klempt}
\affiliation{\BONN}
\affiliation{\JLAB}
%\author{A.V. Anisovich}
%\affiliation{\BONN}
%\affiliation{\PNPI}
\author{V.A. Nikonov}
\affiliation{\BONN}
\affiliation{\PNPI}
\author{A.V. Sarantsev}
\affiliation{\BONN}
\affiliation{\PNPI}

% ...

\collaboration{The CLAS Collaboration}
\noaffiliation

\date{\today}

\begin{abstract}
We report the first measurement of differential and total cross sections 
for the $\gamma d \to K^{0} \Lambda(p)$ reaction, using data 
from the CLAS detector at the Thomas Jefferson National Accelerator Facility.
Data collected during two separate experimental runs were studied with 
photon-energy coverage $0.8-3.6$~GeV and $0.5-2.6$~GeV, respectively.
The two measurements are consistent giving confidence in the method and 
determination of systematic uncertainties. 
The cross sections are compared with predictions from 
the KAON-MAID theoretical model (without kaon exchange),
which deviate from the data at higher $W$ and at forward kaon angles. 
These data, along with previously published cross sections for $K^+\Lambda$ 
photoproduction, provide essential constraints on the nucleon resonance 
spectrum.
A first partial wave analysis has been performed that describes the data 
without the introduction of new resonances.
\end{abstract}

\maketitle

%_____________________________INTRODUCTION_________________________________
\section{Introduction}
New states have been discovered in the spectrum of nucleon resonances
in recent years, which are summarized by the Particle Data Group (PDG) 
\cite{bib:PDG}, in part due to high-precision data from
photon-beam facilities, and also due to theoretical advances in
coupled-channel partial wave analyses \cite{bib:BoGa}.
Some nucleon resonances, or $N^*$'s, have a weak coupling to 
$\pi N$ final states yet may have significant 
branching ratios to final states with strangeness, such as $K^+\Lambda$.
Most of the strangeness photoproduction data comes from reactions
using a proton target. However, protons and neutrons have different 
photocouplings to the $N^*$'s and measurements of cross sections
off the neutron give complementary information \cite{bib:close}. 
Here, we present the first measurements of the reaction 
$\gamma d \to K^0 \Lambda (p)$ where the proton is a spectator. 
(In fact, the proton can contribute in some kinematics through 
final-state interactions, but based on results of other analyses 
we expect these effects to be small here \cite{bib:Zachariou2017}.)
One advantage of studying this reaction is that
the $\Lambda$ is an isosinglet, and hence only $N^*$ resonances 
(and no $\Delta^*$ resonances) can contribute to $s$-channel diagrams, 
thus simplifying the theoretical interpretation of the data.

The measurements are compared with theoretical predictions 
from an approach that is based on a unitarized tree-level Lagrangian model 
\cite{bib:kaonmaid}. The model includes phenomenological couplings 
of $N^*$'s to the $K\Lambda$ final state, based on fits to existing 
kaon production data \cite{bib:g11kplam,bib:g1ckplam,bib:MCNabb}, 
with photocouplings to the $N^*$'s extracted from previous 
measurements (more in Section \ref{sec:model}).
The calculations also include $t$-channel exchange 
based on the Regge model. Since the $K^0$ has no charge or spin, 
the $t$-channel contributions to $K^0 \Lambda$ photoproduction only 
come from an exchange of a strange meson with spin $S=1$, such as a $K^*$.

A comparison between the data and theoretical predictions will allow us 
to obtain information on which $N^*$'s contribute to this reaction. 
In particular, there are many resonances predicted by the constituent 
quark model \cite{bib:isgur,bib:capstickroberts} or by 
lattice gauge theory \cite{bib:edwards} that are not seen in 
experiments and are commonly referred to as ``missing" resonances.
Recent work by the Bonn-Gatchina group \cite{bib:BoGa} has added a 
few new resonances, but many are yet to be observed.

At lower center of mass energies, $W$, only 
$N(1650)1/2^-$, $N(1710)1/2^+$, and $N(1720)3/2^+$ were predicted to 
contribute significantly to $K^{+}\Lambda$ production.
However, the SAPHIR \cite{bib:saphir1,bib:saphir2} and 
CLAS \cite{bib:g11kplam,bib:g1ckplam,bib:MCNabb} photoproduction data 
off a proton target show an enhancement at $W \sim 1.9$~GeV. 
Partial Wave Analyses (PWA) suggested that this corresponds to 
a new resonance, the $N(1900)3/2^+$, which couples only weakly 
to $\pi N$ final states \cite{bib:BoGa}.
Given these findings, data utilizing photoproduction off the neutron are
very important to understand these resonant states.
The measurement of the $\gamma d \to K^0 \Lambda (p)$ cross sections is 
expected to lead to the  determination of excitation coupling strengths, 
relative to the proton.

\begin{comment}
Our new data, for $K^0 \Lambda$ 
photoproduction off the neutron, do not show the same enhancement 
at $W \sim 1.9$~GeV, and suggests that the photocoupling of this 
resonance to the neutron is weak, or perhaps that the enhanced 
cross section at $W \sim 1.9$~GeV in $K^+\Lambda$ photoproduction is 
due to some process other than an intermediate $N^*$ in the $s$-channel. 
\end{comment}
%_____________________________EXPERIMENT_________________________________
\section{The Experiments}
The g10 and g13 datasets were collected using the
CEBAF Large Acceptance Spectrometer (CLAS) at the Thomas Jefferson
National Accelerator Facility.
The experiments used a tagged Bremsstrahlung photon beam \cite{bib:tagger}
created from the primary electron beam of the CEBAF accelerator. These
photons were tagged by determining the scattered electron energy \cite{bib:tagger}.
This allowed tagging photons between 20\% and 95\% of the incident electron 
energy ($E_{0}$) with resolution of $10^{-3}$ $E_{0}$.

\begin{figure}[htbp]
    \centering
      \includegraphics[width=\linewidth, height = 3.0in, keepaspectratio = true]{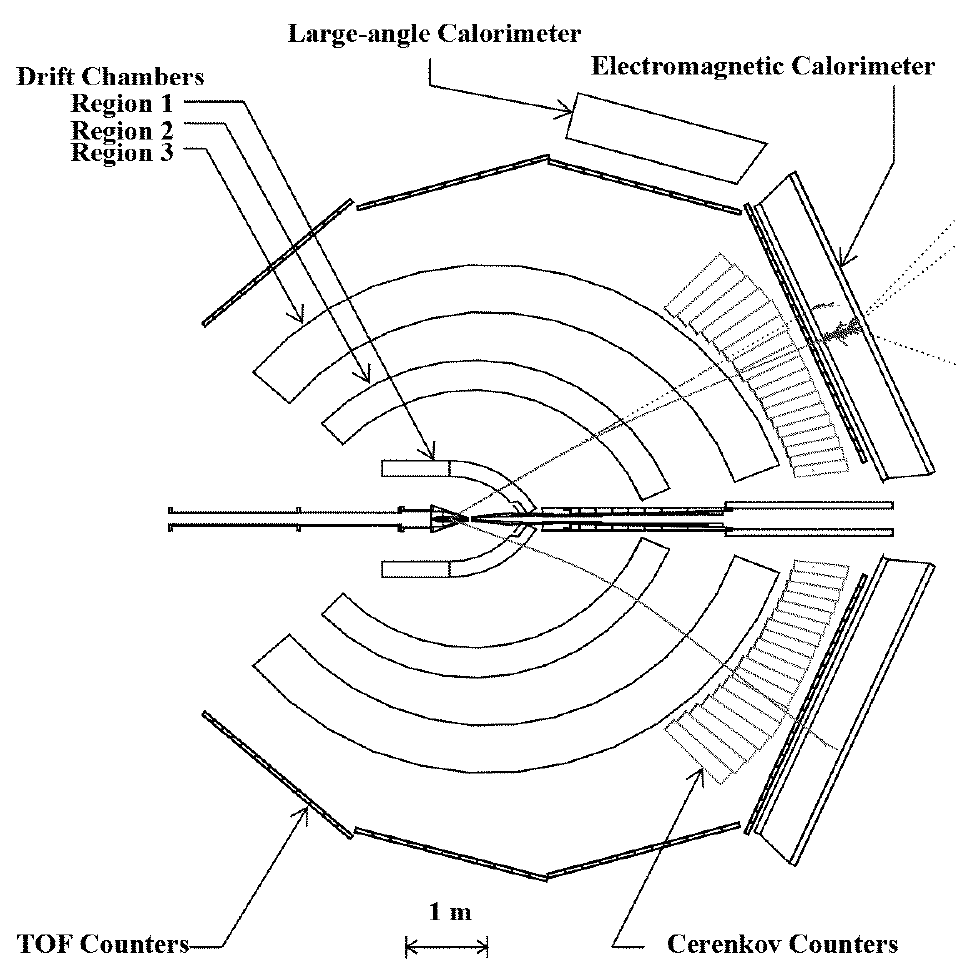}
    \caption{A schematic of CLAS \cite{bib:CLAS} 
(top view, cut along the beamline).}
        \label{fig:CLAS}%
\end{figure}

Some of the generated photons interacted with the liquid deuterium target
and produced a neutral $K^0$ and a $\Lambda$ baryon.
Each of the final state hadrons decayed into pions and
a proton that were tracked by the drift chambers \cite{bib:DC}
in a toroidal magnetic field \cite{bib:CLAS} 
to determine the charge and momenta
of the particles.
The time-of-flight was determined using the
start counter \cite{bib:STg10,bib:STg13}, surrounding the
target, and the (stop) counters on the exterior of CLAS \cite{bib:CLAS}.
A schematic of CLAS can be seen in Fig.~\ref{fig:CLAS}.
The detected particles were then used to reconstruct the momenta and
trajectories of the produced kaon and $\Lambda$ in the offline analysis.
%insert schematic of CLAS

\subsection{g10 Experiment}
%The unpolarized photon beam for the g10 experiment was created by colliding an
%electron beam into a gold foil that caused a range of photon energies
%to be sent down the beam pipe to a unpolarized liquid-deuterium target.
The g10 experiment directed the CEBAF electron beam onto a gold foil to 
produce an unpolarized bremsstrahlung photon beam, which then struck an 
unpolarized liquid-deuterium target.
The target chamber was conical, measuring 24~cm in length with a maximal diameter of 4~cm.
The center of the target was positioned 25~cm upstream
from the CLAS center.
%The energy of the photons emitted from the foil were calculated by
%identifying the corresponding electron in the
%tagging spectrometer \cite{bib:tagger}.
For this experiment the incident electron energy was $E_{e−} = $ 3.767~GeV,
which allowed a maximum tagged photon energy of about 3.6~GeV.
%The outgoing electron traveled through a magnetic field and into the tagger
%that separated the electrons based on energy.

The analysis on this dataset was limited to photon energies
between 1.0~GeV and 3.0~GeV where event rates were the largest.
%The g10 run period used the detection of two charged particles in
%different TOF sectors along with a hit in the electromagnetic
%calorimeter \cite{bib:EC} as a trigger.
The torus had two different current settings, +2250~A and +3375~A 
\cite{bib:g10ppim}.
%The magnetic current in this run period had two different settings.
%The first part had the torus magnet set at +2250 Amps,
%while the second part was set at +3375 Amps \cite{bib:g10ppim}.
Each magnet setting was kept for roughly half of the g10 beam time.
The positive polarity, which bends negatively charged particles towards 
the beamline, combined with the high torus setting, resulted in some
low-momentum $\pi^{-}$ tracks curling far enough inward to never be seen 
by the time-of-flight (stop) counters.
Therefore, this analysis only investigated the data set with the
torus magnet set at +2250~A as to increase the probability of
detecting lower momentum $\pi^{-}$'s.
%This setting was chosen in the analysis as the lower field strength increases the probability of
%detecting low momentum $\pi^{-}$'s.

%In both torus settings, a  positive polarity was chosen such that negatively charged pions were bent towards the beamline, resulting in the possibility of a
%low momentum ($<150$ MeV/c) $\pi^{-}$ to be lost
%before reaching the outer TOF wall.

\subsection{g13 Experiment}
%The g13 experiment was carried out with both circularly and linearly polarized
%photon beams and with multiple beam energies.
This analysis used the g13 experiment's data 
with circularly polarized photons that 
were generated using a polarized electron beam at an energy of 2.65~GeV.
The torus magnet current was set to $-1497$~A to have larger efficiency for 
low momentum $\pi^{-}$'s that bent away from the beamline, in contrast to 
g10's positive torus polarity.
%The g13 experiment was split into two parts,
%g13a and g13b, allowing the collection of data using a circular and linear photon polarization, respectively \cite{bib:g13proposal}.
%Both used specialized radiators to transform the electron beam
%into a polarized photon beam.
%The collimation needed for enhancing the linear polarization made a cross section analysis for g13b difficult, and therefore was not used in this study.
%The g13a data was characterized by high luminosity as the experiment was designed to measure cross sections precisely.
%Its circularly polarized photon beam was produced from the interaction
%of a longitudinally polarized electron beam with a gold foil.
%The g13b experiment used a diamond target with tight collimation
%to produce a linearly polarized photon beam.
A conical 40-cm-long unpolarized liquid-deuterium target was used during 
the g13 experiment.
This was positioned 20~cm upstream from the CLAS center with a maximal diameter 
of 4~cm.
%A current of $-1497$ A was used in the torus magnet to bend
%negatively charged particles away from the beamline,
%in contrast to g10's positive polarity.
This set-up was intended to maximize the acceptance of low-momentum
$\pi^{-}$'s that resulted from the decays of hyperons.
%The larger subset of g13a data was collected with incident electron energy of 2.65 GeV.
These data were used for the cross section determination presented here 
due to its large energy overlap with the g10 data set.

%The g13a portion of the experiment was in itself broken up into two parts.
%The first part was 3-pass data (three passes through the accelerator loop).
%These data utilized an incident electron beam of
%1.996 GeV to create the photon beam.
%The second part was 4-pass data, utilizing a beam energy of 2.65 GeV.
%The 4-pass data are used in this analysis to provide a measurement with higher
%statistics.
% complimentary to that of the g10 measurement.
%The 3-pass data had a luminosity of about 68\% of the 4-pass data.
%Due to the difference in electron beam energies, different tagger
%bins are associated with the same photon energy depending on run number.

%_____________________________DATA_ANALYSIS______________________________
\section{Data Analysis}
\begin{figure*}[!t]
    \centering
    \includegraphics[width= 0.33\textwidth, height = 2 in, keepaspectratio = true]{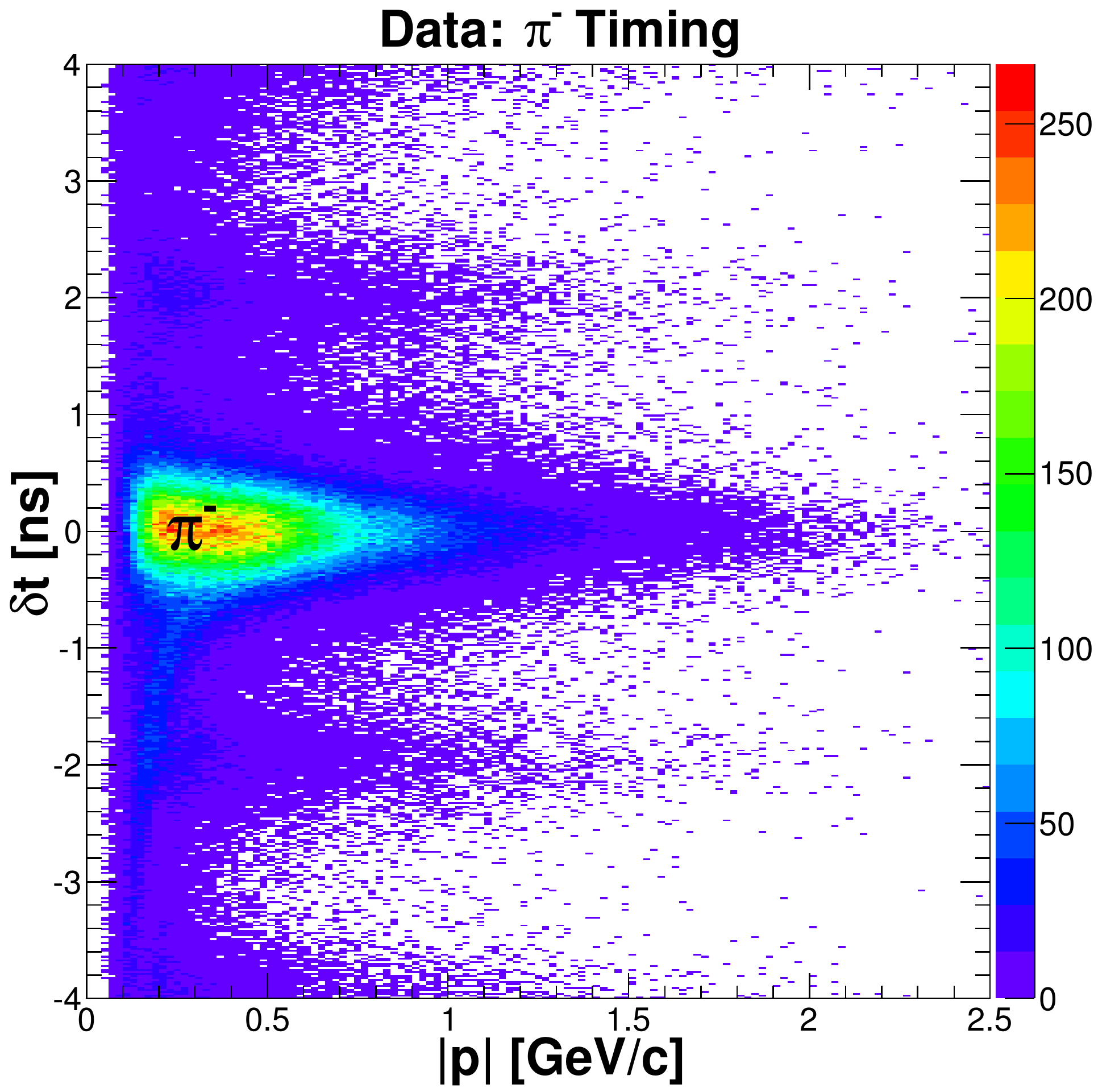}
    \includegraphics[width= 0.33\textwidth, height = 2 in, keepaspectratio = true]{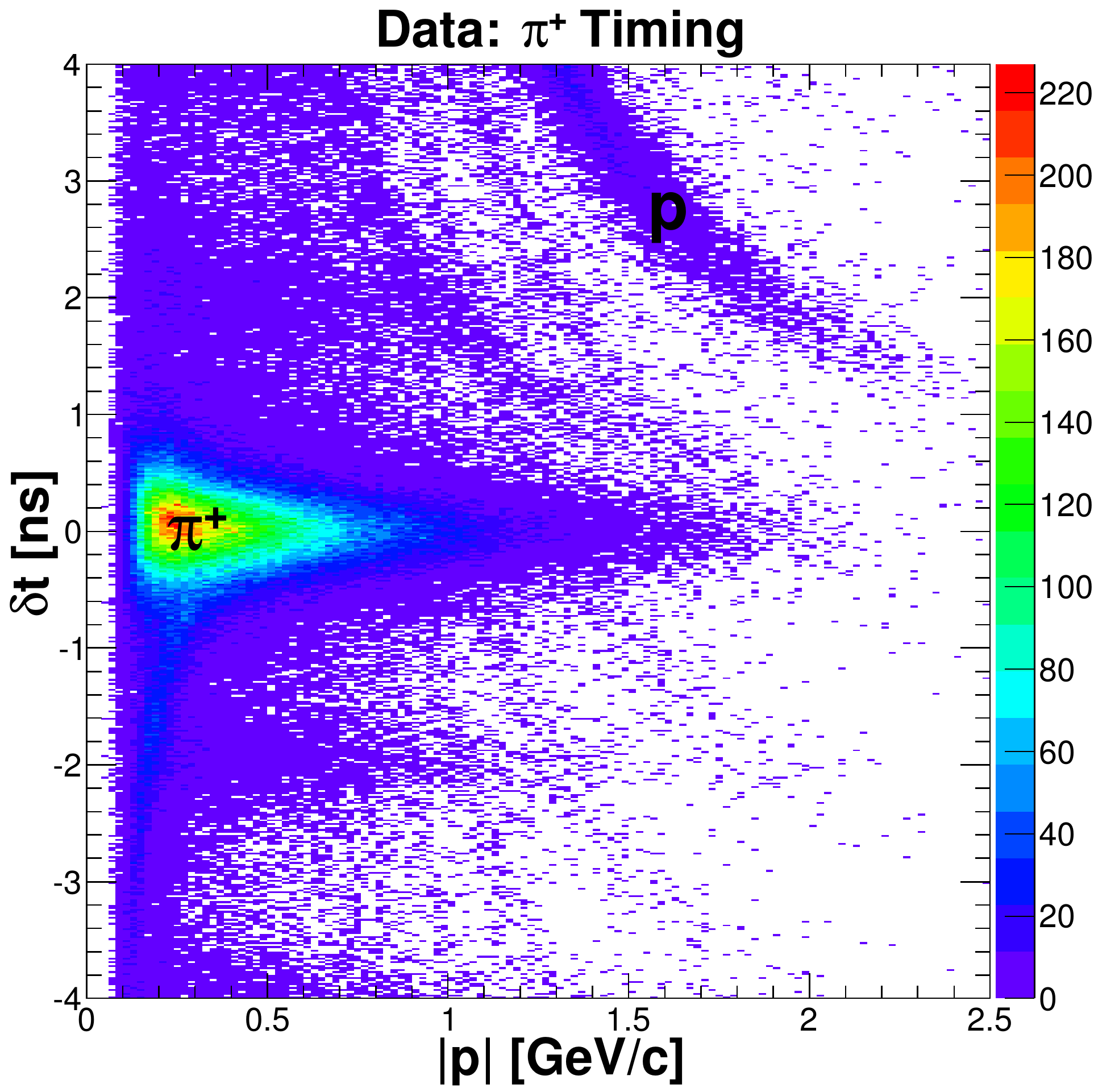}
    \includegraphics[width= 0.33\textwidth, height = 2 in, keepaspectratio = true]{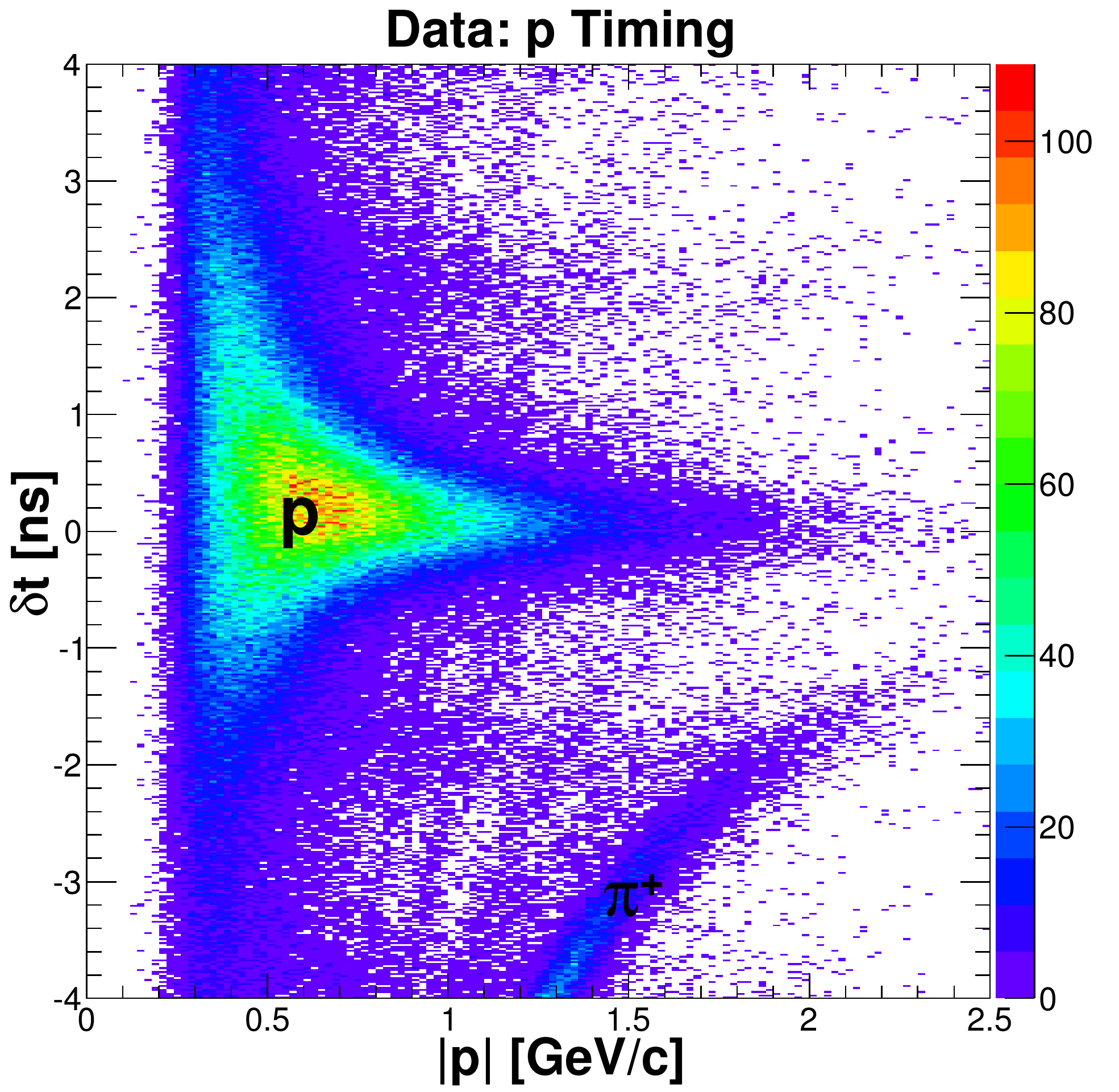}
    \caption{A very small subset of the raw g13 event tracks before any cuts. The tracks shown differ only by charge and assumed mass. This figure represents (left) negatively charged tracks with an assumed mass of $m_{\pi}$, (middle) positively charged tracks with an assumed mass of $m_{\pi}$, and (right) positively charged tracks with an assumed mass of $m_{p}$.}
    \label{fig:deltaTmom}
\end{figure*}
The different run conditions of the g10 and g13 experiments allowed a check
on the reproducibility of this first
$K^{0}\Lambda$ cross section measurement.
Differences in these independent measurements
include the photon tagger energy range, photon flux, and
torus field strength and polarity.
\begin{comment}
\begin{figure*}[!t]
    \centering
    \begin{subfigure}[t]{0.3\textwidth}
        \centering
        \includegraphics[width= \textwidth, height = 2 in, keepaspectratio = true]{pimtimeEXAMPLE}
        \caption{}
        \label{fig:pimtimeEXAMPLE}
    \end{subfigure}
    ~
    \begin{subfigure}[t]{0.3\textwidth}
        \centering
        \includegraphics[width= \textwidth, height = 2 in, keepaspectratio = true]{piptimeEXAMPLE}
        \caption{}
        \label{fig:piptimeEXAMPLE}
    \end{subfigure}
    ~
    \begin{subfigure}[t]{0.3\textwidth}
        \includegraphics[width= \textwidth, height = 2 in, keepaspectratio = true]{prtimeEXAMPLE}
        \caption{}
        \label{fig:prtimeEXAMPLE}
    \end{subfigure}
    \caption{Time difference for a small subset of raw events before for 
(a) negatively charged tracks with an assumed $\pi^-$ mass, 
(b) positively charged tracks with an assumed $\pi^+$ mass, and 
(c) positively charged tracks with an assumed proton mass. 
The low-count background stripes are due to early/late triggers and 
show the 2~ns time structure of the beam.}
    \label{fig:deltaTmom}
\end{figure*}
\end{comment}

\subsection{Particle Identification}
The short lifetime and neutral charge of the reaction products
of interest, $K^{0}$ and $\Lambda$, make their direct detection
virtually impossible.
The particles were reconstructed through their decays:
$K^{0}\rightarrow K^{0}_{S}\rightarrow \pi^{+}\pi^{-}$ and
$\Lambda\rightarrow p\pi^{-}$.
Having no particles detected directly from the reaction vertex
required an extra step to determine the decay vertex, which was used to account
for energy losses due to ionization (and momentum corrections).
These corrections are essential for
making a direct and reliable comparison of
data and simulation.

The final-state particles representing the reaction of interest are three pions and a proton.
%This made the identification of the measured decay
%products relatively straightforward.  
Particle identification consisted of a comparison between the
measured time-of-flight, $t_{m}$,
and the calculated time using the particle's assumed mass and momentum (as extracted from tracking); 
\begin{equation} 
\delta t=t_{m}-\frac{D}{\beta c}=t_{m}-\ \frac{D\sqrt{p^2+m^2}}{pc},
\label{EQ1}               
\end{equation} 
where $D$ is the reconstructed path length of the particle 
from the event vertex to the TOF counters, 
$m$ is the assumed mass of the particle, and
$t_{m}$ is the time-of-flight as calculated by taking
the difference between the TOF time
and the event start time.
Particle identification was performed separately
for positive and negative tracks.
Fig.~\ref{fig:deltaTmom} shows a very small subset of the raw g13 data,
in which $\delta t$ is determined for each track, given its measured
quantities ($t_{m}$, charge, and momentum) for assumed masses of a $\pi$ and $p$.
The time difference about $\delta t =0$ was fit as a function of
momentum with a Gaussian for several momentum bins;
a $2\sigma$ ($3\sigma$) cut about the centroid of $\delta t$ was used to
identify particles in g13 (g10) data. Fig.~\ref{fig:deltaTmom} shows 
horizontal bands at $\delta t = \pm 2$~ns and $\pm 4$~ns that reflect the
2~ns RF period of the CEBAF electron beam.
%This resulted in a fit about these Gaussian centroids
%as well as a fit to the Gaussian sigmas. 

%\onecolumngrid

%\twocolumngrid
%\FloatBarrier
\subsection{Event Selection}
Once the candidate events with all the required particles were identified,
their tracks were paired to reconstruct the possible
$K^{0}_{S}$ and $\Lambda$ particles.
The $K^{0}_{S}$ decays 69\% of the time into a
$\pi^{+}\pi^{-}$ pair \cite{bib:PDG},
while the $\Lambda$ has a 64\% branching ratio to the $\pi^{-}p$
channel \cite{bib:PDG}.
It cannot be certain {\it apriori} which of the two $\pi^{-}$'s was the
partner of the proton and which one of the $\pi^{+}$,
so each combination was considered.
The $\pi^{-}p$ pair that yielded an invariant mass
closest to the $\Lambda$ mass was chosen.
From both simulation and data studies,
it was shown that less than 1.0\% of
surviving events were then paired incorrectly \cite{bib:g10note,bib:g13note}.
This showed that each $\pi^{-}$
could be reliably assigned to a corresponding $p$
or $\pi^{+}$ (when a $K^{0}_{S}\Lambda$  event existed)
and was used for $K^{0}_{S}$ and $\Lambda$
reconstruction in this analysis.

Several corrections and cuts were applied before the
final yield extraction was done.
The momenta of the tracks was 
corrected for the energy lost as the particles
passed through the target and start counter \cite{bib:ELOSS}.
Slight corrections were also necessary for the momentum of each track,
due to uncertainties in the magnetic field,
%transition through the drift chambers due to inefficiencies
and for the tagged photon energy, caused by 
the sag of the tagger focal plane \cite{bib:g11kplam}.
Cuts were also made to remove poorly performing tagger counters and time-of-flight paddles.
Events associated with beam trips were also cut from the final analysis.
%\cite{bib:g13studies}.

Every particle that traverses through CLAS can be described by its
production vertex, momentum ($\vec{p}$), and mass.
To increase reliability, all tracks that were reconstructed close to the edges of the
detector were removed from both data and simulation \cite{bib:g11kplam}. 
These trajectories were identified based on the decrease in the number of
reconstructed particles in finite bins of the vertex, momentum ($\vec{p}$), and mass.
%CLAS itself has blind spots, which were modeled as momentum
%dependent trajectories in $\theta$ and $\phi$.
These fiducial cuts change with each experiment, due to different magnetic fields and target locations.
%or cite williams thesis

\begin{figure}[htbp]
    \centering
    \includegraphics[width= 0.49\linewidth, height = 2.5 in, keepaspectratio = true]{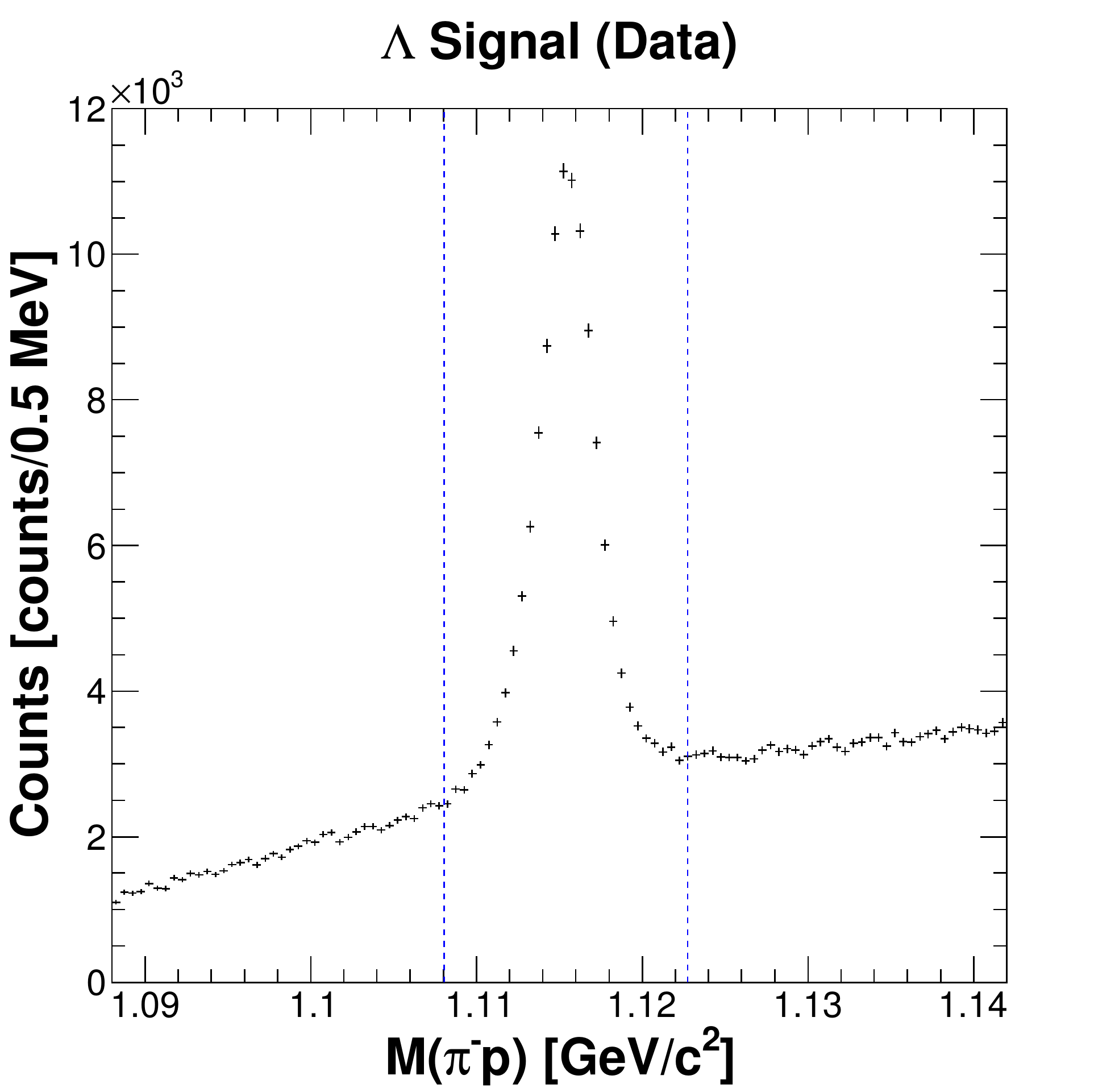}
    \includegraphics[width= 0.49\linewidth, height = 2.5 in, keepaspectratio = true]{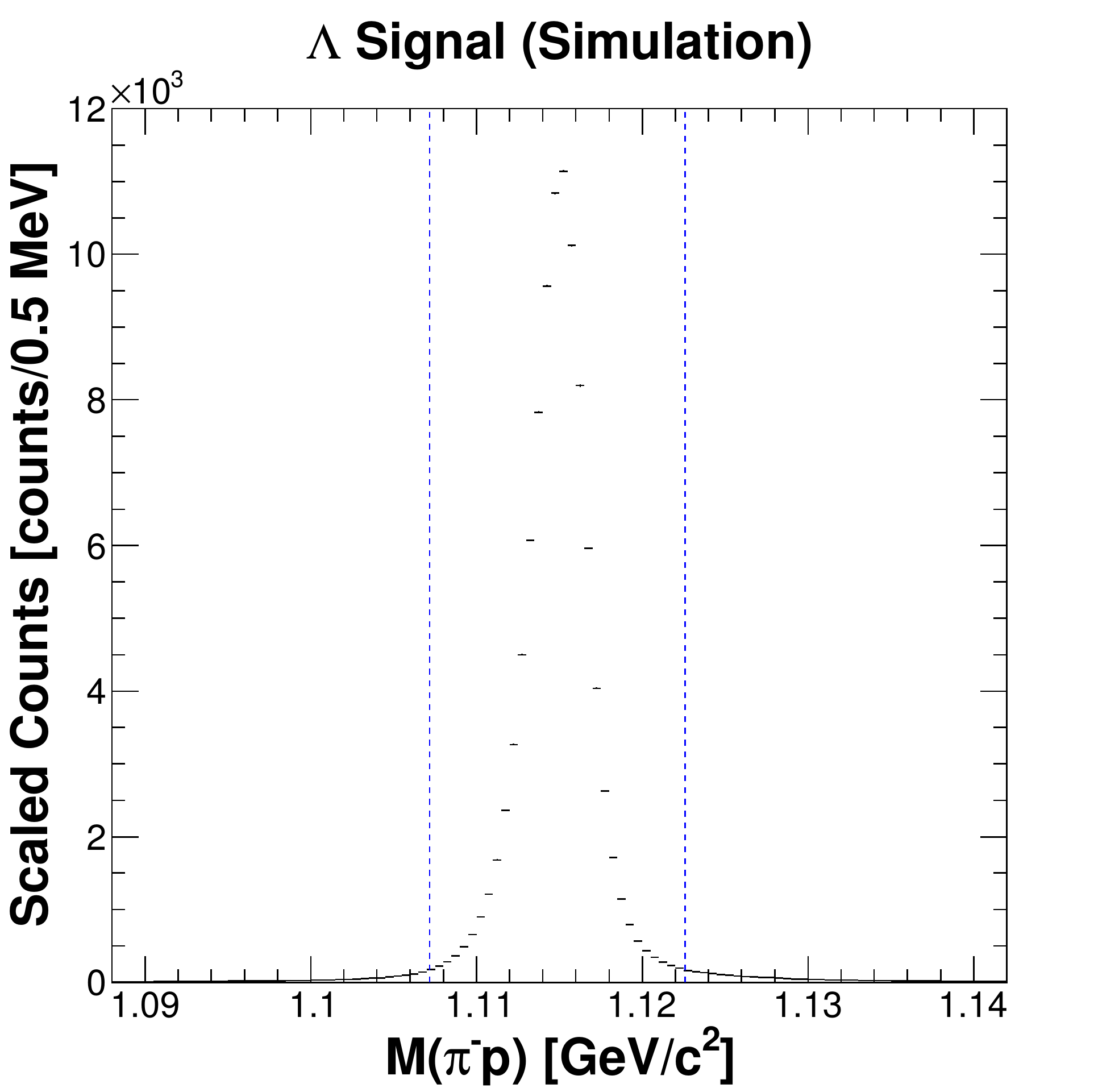}
    \caption{The invariant mass of the $\pi^{-}p$ pair for both (left) data and (right) simulation. %The solid curves represent a first order polynomial along with a Gaussian fit, allowing the extraction of a centroid and $\sigma$.
    The dotted lines represent 4$\sigma$ from the centroid, where $\sigma\approx 2\ \text{MeV}/\text{c}^{2}$.}
    \label{fig:lambdafit}
\end{figure}

\begin{figure}[htbp]
    \centering
    \includegraphics[width= 0.49\linewidth, height = 2.5 in, keepaspectratio = true]{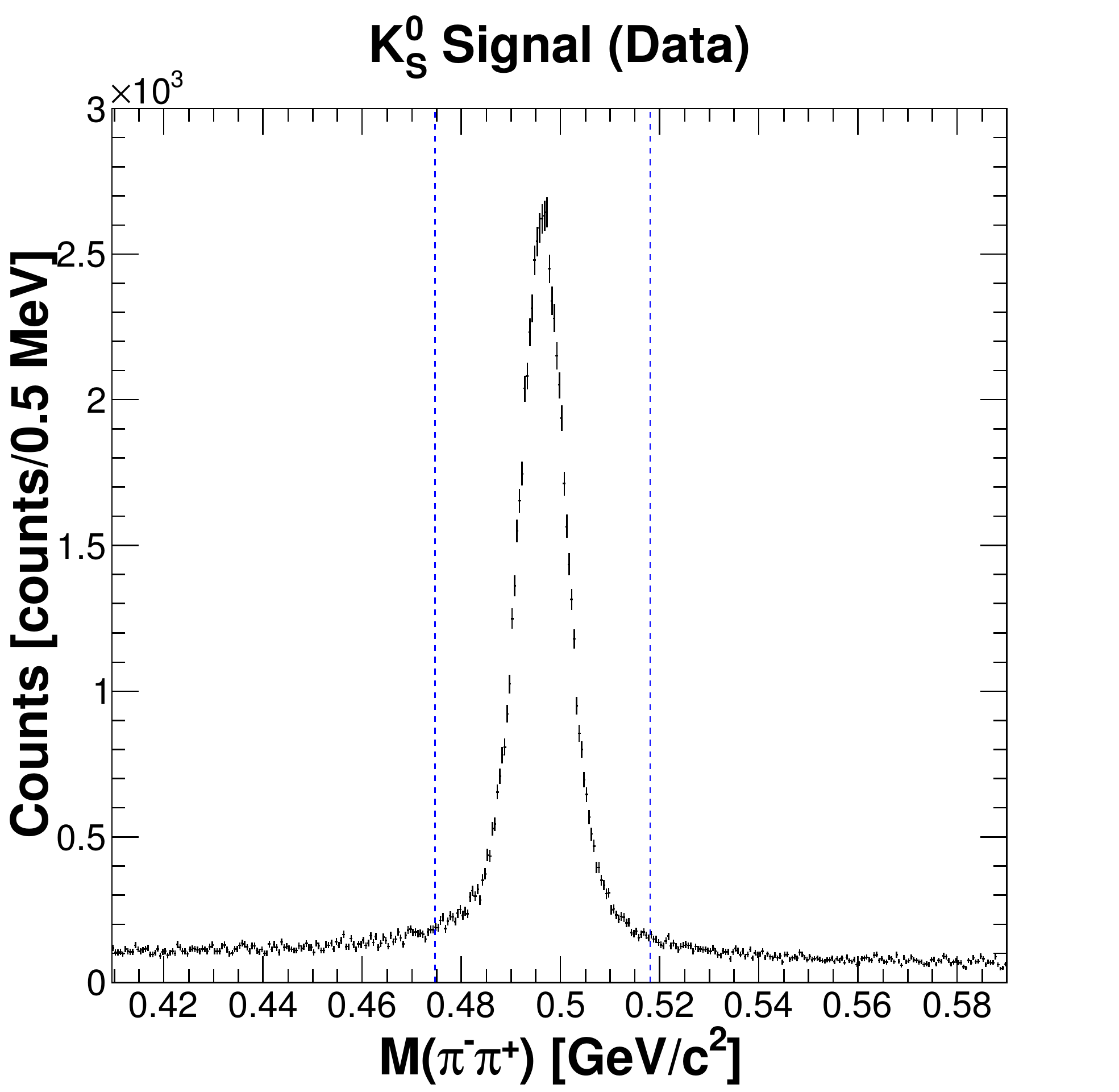}
    \includegraphics[width= 0.49\linewidth, height = 2.5 in, keepaspectratio = true]{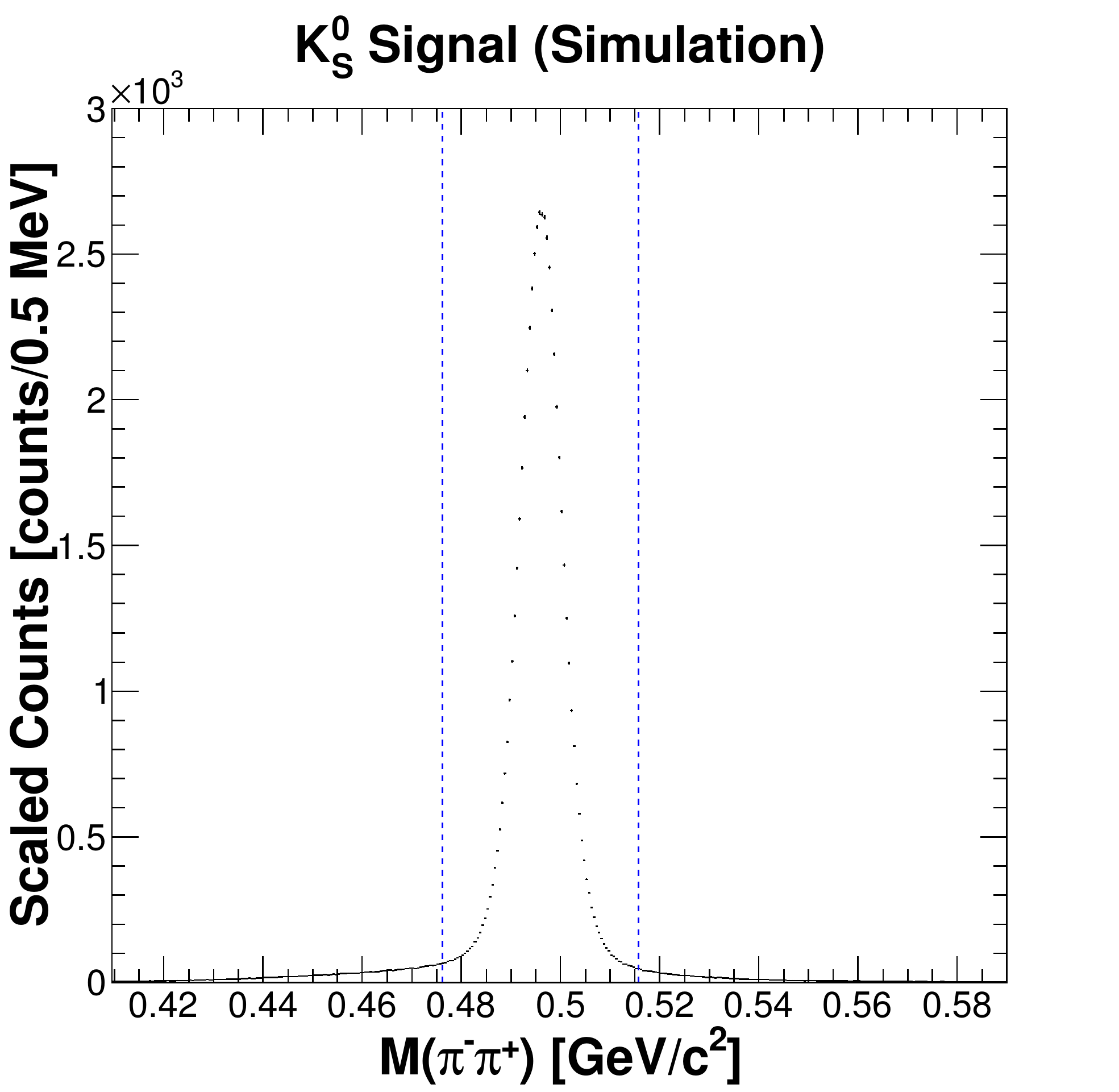}
    \caption{The invariant mass of the $\pi^{+}\pi^{-}$ pair after a 4$\sigma$ cut on $M(\pi^{-}p)$ for both (left) data and (right) simulation. %The solid curves represent a first order polynomial along with a Gaussian fit, allowing the extraction of a centroid and $\sigma$.
    The dotted lines represent 4$\sigma$ from the centroid, where $\sigma\approx 5\ \text{MeV}/\text{c}^{2}$.%It should be noted that the Kaon signal is two to three times wider than the $\Lambda$ signal.
    }
    \label{fig:kaonfit}
\end{figure}

Figures \ref{fig:lambdafit} and \ref{fig:kaonfit} show the
reconstructed invariant mass distributions of $\pi^{+}\pi^{-}$
and $\pi^{-}p$, respectively.
One can clearly see the $K^{0}_{S}$ and $\Lambda$ peaks.
The peaks sit on top of background, which was mostly due
to non-resonant $p\pi^{+}\pi^{-}\pi^{-}X$ production.
The phase space background can be reduced by a cut on
the opposing particle's ($K^{0}_{S}$ or $\Lambda$) mass distribution
(a 4$\sigma$ cut was used in this analysis).
To illustrate that the data has peaks where they are expected,
a simulation of $\gamma d\rightarrow K^{0}_{S}\Lambda(p)$
was compared with the data.
At this point the data contained a large amount of background.
To reduce this background, cuts on the invariant mass
(as discussed above) were imposed on the data and simulation.
The peak location, width of these signal peaks,
and a representation of where a 4$\sigma$ cut would lie is
shown in Fig.~\ref{fig:lambdafit} and \ref{fig:kaonfit}
for the reconstructed $\Lambda$ and $K^{0}_{S}$, respectively.
%Fits used on the data and simulation can be seen in Figures \ref{fig:lambdafit} and \ref{fig:kaonfit}.
\begin{comment}
\begin{figure}[htbp]
    \centering
    \begin{subfigure}[t]{0.48\linewidth}
        \includegraphics[width= \linewidth, height = 2.5 in, keepaspectratio = true]{DataLamFit}
        \caption{%The data centroid was extracted to be 1.1155 GeV and the corresponding $\sigma$ to be $1.73\times 10^{-3}$ GeV.
        }
        \label{fig:DataLamFit}
    \end{subfigure}
    \begin{subfigure}[t]{0.48\linewidth}
        \includegraphics[width= \linewidth, height = 2.5 in, keepaspectratio = true]{SimLamFit}
        \caption{%The Simulation centroid was extracted to be 1.1148 GeV and the corresponding $\sigma$ to be $1.80\times 10^{-3}$ GeV.
        }
        \label{fig:SimLamFit}
    \end{subfigure}
    \caption{Shown is the invariant mass of the $\pi^{-}p$ pair for both the (a) data and (b) simulation. %The solid curves represent a first order polynomial along with a Gaussian fit, allowing the extraction of a centroid and $\sigma$.
    The dotted lines represent 4$\sigma$ from the centroid, where $\sigma\approx 2\ \text{MeV}/\text{c}^{2}$.}
    \label{fig:lambdafit}
\end{figure}
\end{comment}

\begin{comment}
\begin{figure}[htbp]
    \centering
    \begin{subfigure}[t]{0.48\linewidth}
        \includegraphics[width= \linewidth, height = 2.5 in, keepaspectratio = true]{DataKaonFit}
        \caption{%The data centroid was extracted to be 0.4964 GeV and the corresponding $\sigma$ to be $5.44\times 10^{-3}$ GeV.
        }
        \label{fig:DataKaonFit}
    \end{subfigure}
    ~
    \begin{subfigure}[t]{0.48\linewidth}
        \includegraphics[width= \linewidth, height = 2.5 in, keepaspectratio = true]{SimKaonFit}
        \caption{%The Simulation centroid was extracted to be 0.4958 GeV and the corresponding $\sigma$ to be $4.89\times 10^{-3}$ GeV.
        }
        \label{fig:SimKaonFit}
    \end{subfigure}
    \caption{Shown is the invariant mass of the $\pi^{+}\pi^{-}$ pair after a 4$\sigma$ cut on $M(\pi^{-}p)$ for both the (a) data and (b) simulation. %The solid curves represent a first order polynomial along with a Gaussian fit, allowing the extraction of a centroid and $\sigma$.
    The dotted lines represent 4$\sigma$ from the centroid, where $\sigma\approx 5\ \text{MeV}/\text{c}^{2}$.%It should be noted that the Kaon signal is two to three times wider than the $\Lambda$ signal.
    }
    \label{fig:kaonfit}
\end{figure}
\end{comment}

\subsection{Yield Extraction}
\begin{figure}[htbp]
    \centering
      \includegraphics[width=\linewidth, height = 3.0in, keepaspectratio = true]{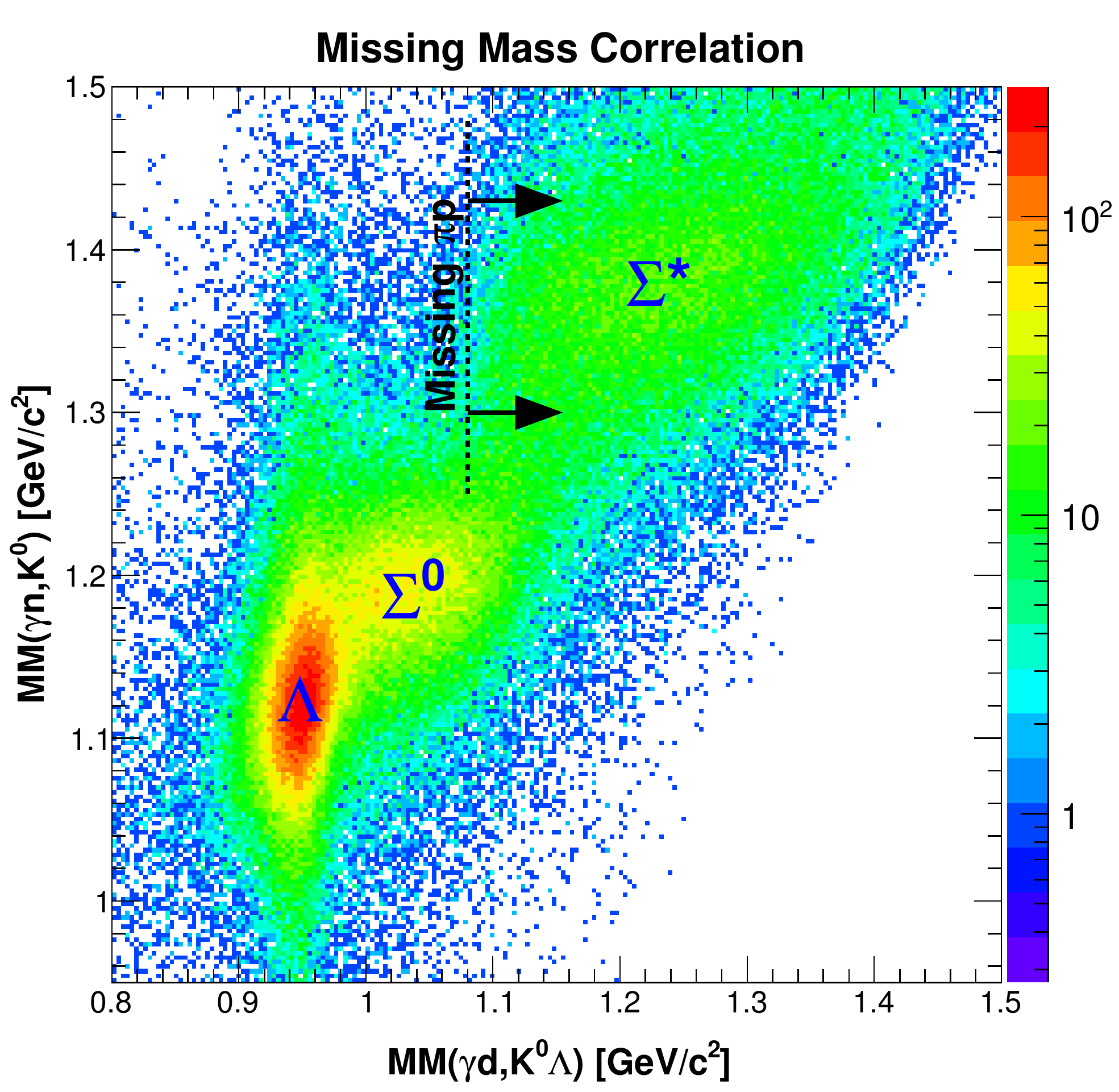}
    \caption{(g13 data) The missing mass phase space for $MM(\gamma n,\pi^{+}\pi^{-})$ versus $MM(\gamma d,\pi^{+}\pi^{-}\pi^{-}p)$. Clear signals can be seen between the expected hyperons in this reaction. The lower left peak corresponds to the $K^{0}_{S}\Lambda$ channel, while the other peaks seen correspond to the background channels $K^{0}_{S}\Sigma^{0}$ and $K^{0}_{S}\Sigma^{*}$, from left to right.
    %Unfortunately, the $\Lambda$ and $\Sigma^{0}$ distributions have significant overlap. Fitting is required to separate the yields, rather than an event by event sorting. Events with an extra pion, including $\Sigma^{*}$, clearly do not have a significant contribution to the missing proton peak seen in $MM(\gamma d,\pi^{+}\pi^{-}\pi^{-}p)$.
    The edge of the missing pion and proton peak is indicated by the black dotted line drawn at $MM(\gamma d,\pi^{+}\pi^{-}\pi^{-}p) = 1.08$~GeV/c$^{2}$.}
        \label{fig:missmass2d}%
\end{figure}

Extraction of the exclusive $\gamma d \rightarrow K^{0}_{S}\Lambda(p)$ events from the sample of $\gamma d \rightarrow \pi^{+}\pi^{-}\pi^{-}pX$ events requires the background contributions to be identified and removed (or accounted for).
Also, final-state-interaction events need to be eliminated or strongly suppressed.
Previous studies of the reaction of interest \cite{bib:g13proposal} have shown that the distribution of the missing mass off the kaon, $MM(\gamma n,K^{0}_{S})$ (where $n$ was assumed to be at rest), versus the missing mass off $K^{0}_{S}\Lambda$, $MM(\gamma d,K^{0}_{S}\Lambda)$, was useful in understanding background contributions from reactions with higher-mass hyperons such as $\Sigma^{0}$ and $\Sigma^{*}$.
This can be seen in Fig.~\ref{fig:missmass2d}.

The events of interest yield a peak in $MM(\gamma n,K^{0}_{S})$ at the $\Lambda$ mass.
This peak was much wider, compared to $K^{+}\Lambda$ production off the free proton, since the Fermi momentum of the target neutron was not taken into account in the calculation of $MM(\gamma n,K^{0}_{S})$.
This quantity, due to the undetected nucleon, was not sufficient to remove background.
While the $\Sigma^{0}$ cannot be removed with a simple cut, the $\Sigma^{*}$ contributions can be 
reduced to a negligible amount by removing all events with $MM(\gamma d,K^{0}_{S}\Lambda) > 1.05$~GeV/c$^{2}$.
This means the $\Sigma^{*}$ signal does not extend underneath the $K^{0}_{S}\Lambda$ peak when working with the projection onto $MM(\gamma d,K^{0}_{S}\Lambda)$.
A similar argument was made for $K^{*}$, or other events with a missing pion.
Therefore, $MM(\gamma d,K^{0}_{S}\Lambda)$ was used for the yield extraction as discussed throughout this document.

\begin{figure}[htbp]
    \centering
      \includegraphics[width=\linewidth, height = 3in, keepaspectratio = true]{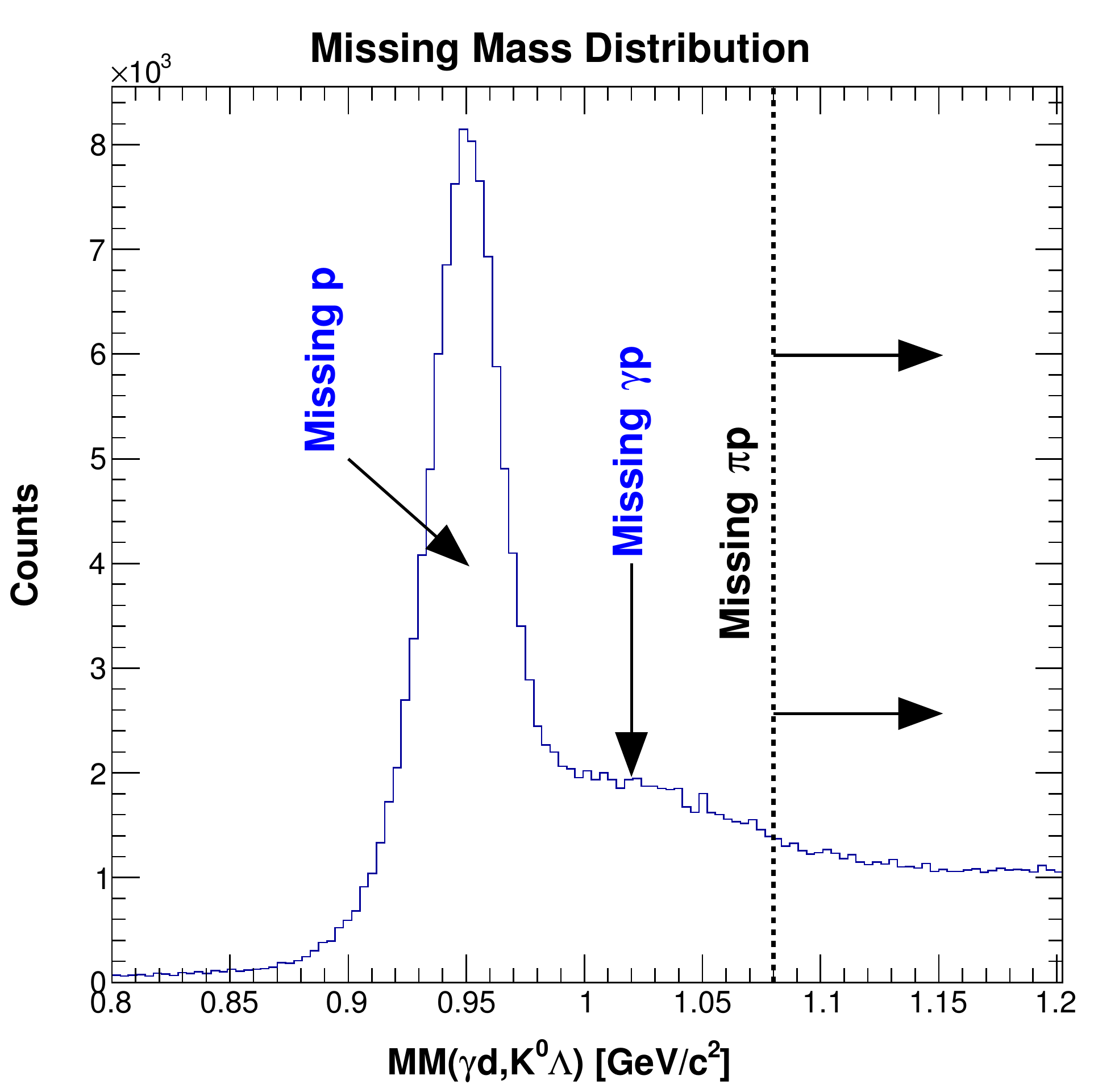}
    \caption{The missing mass distribution peaks about the proton mass due to the creation of $K^{0}_{S}\Lambda$. Another signal can be seen from a missing proton plus photon that was largely associated with the $K^{0}_{S}\Sigma^{0}$.
    %Events with an extra pion including $\Sigma^{*}$ do not contribute to peak associated with $K^{0}_{S}\Lambda$, even if they survived the cuts used in isolating the signal. This was because
    An extra missing pion would push the missing mass to the right of the black dotted line (drawn at the same location as in Fig.~\ref{fig:missmass2d}).}
        \label{fig:missmass1D}%
\end{figure}

The distributions in Fig.~\ref{fig:missmass1D} illustrate the missing mass after cuts on the invariant mass of the $\Lambda$ and $K^{0}_{S}$.
Although much background remains, it is clear where the corresponding signals from the $\gamma d\rightarrow K^{0}_{S}\Lambda(p)$ and $\gamma d \rightarrow K^{0}_{S}\Sigma^{0}p\rightarrow K^{0}_{S}\Lambda (\gamma p)$ reactions are located.
The later cuts on missing mass and missing momentum remove any significant contribution from events associated with the production of an extra $\pi^{0}$ (or $\pi^{+}$) such as in the case of the higher-mass hyperons $\Sigma^{*}$ and $\Lambda^{*}$ or higher-mass kaons.
The yield for $\gamma d \rightarrow K^{0}_{S}\Lambda(p)$ can be determined by fitting the missing proton peak after the analysis cuts.
To do this the $K^{0}_{S}\Sigma^{0}$ background shape must be understood.
Fitting the full spectrum of both the proton peak (corresponding to the missing mass of the $K^{0}_{S}\Lambda$) and the proton plus photon peak ($K^{0}_{S}\Sigma^{0}$) was problematic due to the overlap of these signals.

%this figure definately needs to be updated
%higher quality
%no cut on missing mass, or if there is that line should be shown on previous figure. But this is not used in the final yield extraction, so that should be made abuntantly clear.

To extract a more reliable yield, the fitting of $K^{0}_{S}\Sigma^{0}$ was approached by means of only describing the leading edge of the distribution.
Generated data allowed a very good approximation of background contributions, and these were used to perform background subtraction as described in the next section.
Specifically, the shape of the background was determined by fitting the simulated $K^{0}_{S}\Sigma^{0}$ spectrum after it was processed through the modeled detector.
This shape was then scaled to match the distribution of our actual data.
The yields were extracted by the integration of the signal and by scaling the background shapes to the data.

\subsection{Background}
The reaction of interest was $\gamma d\rightarrow K^{0}\Lambda (p)$.
To measure this process the decay products of the $K^{0}$ and $\Lambda$ were detected.
Therefore the final state particles that were detected were $\pi^{-}\pi^{+}\pi^{-}p$.
The four tracks could be produced several different ways.
The backgrounds can be attributed to two categories.
The first category was a five (or more) track background, where one (or more) tracks were missed by CLAS.
The second category of background processes was from a four track background.

\subsubsection{Hyperon Backgrounds} \label{sec:fiveparticleback}
By extracting the yield through the missing mass, it was likely that any process producing an extra pion (or other massive particle) was well separated from the spectator proton missing mass measured by $MM(K^{0}\Lambda) = \sqrt{(P_{d}+P_{\gamma} -P_{K^{0}} - P_{\Lambda})^{2}}$, where $P_{i}$ is the four-momentum of the given particle.
Near the missing mass signal the most prevalent five track background was identified as $\gamma d \rightarrow K^{0}\Sigma^{0}p \rightarrow K^{0}_{S}\Lambda(\gamma p)\rightarrow \pi^{-}\pi^{+}\pi^{-}p(\gamma p)$.
Nonetheless other background channels were also explored.

\begin{figure}[htbp]
    \centering
    \includegraphics[width= 0.9\linewidth, height = 2.5 in, keepaspectratio = true]{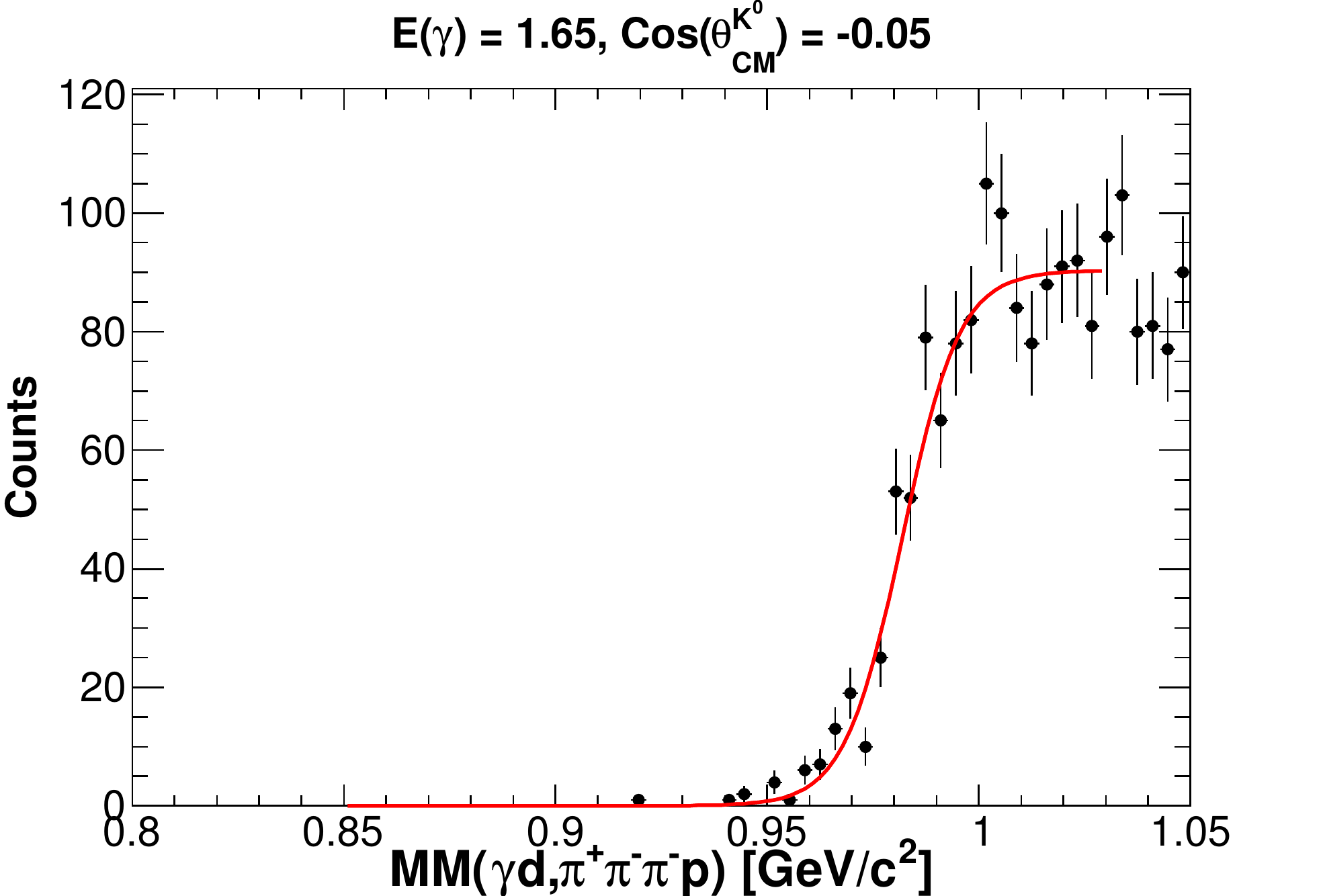}

    \includegraphics[width= 0.9\linewidth, height = 2.5 in, keepaspectratio = true]{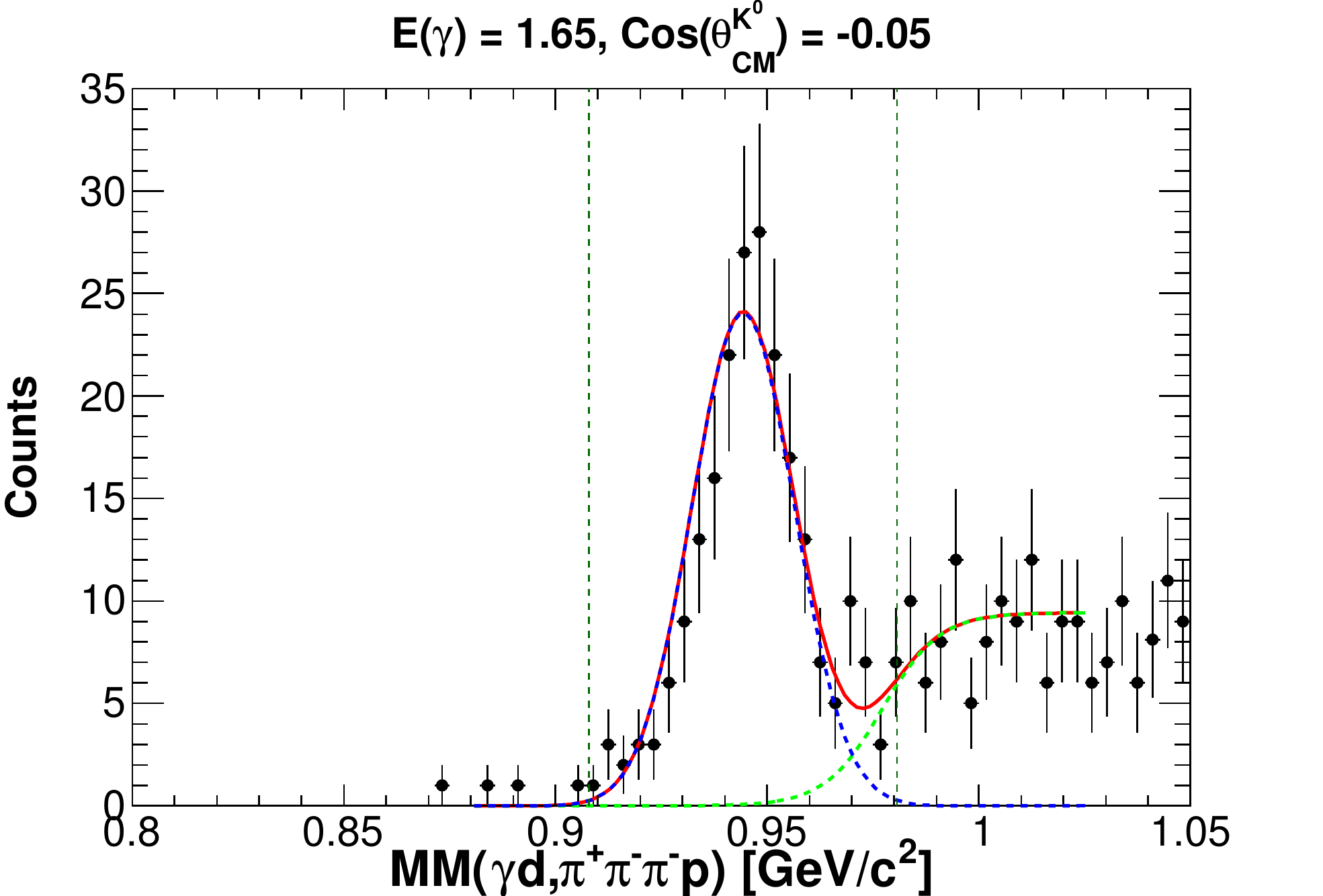}
    \caption{The top panel is a hyperbolic tangent curve fit to the missing 
mass of $\gamma d \to K^{0}_{S}\Sigma^{0}p \to \pi^{-}\pi^{+}\pi^{-}p(\gamma p)$ 
from simulation. 
The bottom panel is a fit to the missing mass of an example data bin. 
The fit uses a Gaussian and a hyperbolic tangent shape that was parameterized 
from the simulation. This is an example of the g13 data fit to extract 
the number of events that were missing only one proton.}
    \label{fig:yieldfit}
\end{figure}

%\subsubsection{$\gamma d \rightarrow K^{0}\Sigma^{0}(p)$}
The $K^{0}_{S}\Sigma^{0}$ background could not be separated from the $K^{0}_{S}\Lambda$ signal except through the missing mass, as this still produced a peak at the $\Lambda$ and $K^{0}_{S}$ invariant masses.
The characteristic shape of this background was explored through simulation.
When extracting the yield for the $K^{0}_{S}\Lambda$ channel, a fit to this background shape was used to subtract the $K^{0}_{S}\Sigma^{0}$ events, which can be seen in Fig.~\ref{fig:yieldfit}.
Simulation showed that the edge of the $K^{0}_{S}\Sigma^{0}$ distribution consistently resulted in a sigmoidal shape.
Several fitting functions (with sigmoidal properties) proved reasonably consistent, yet the hyperbolic tangent function proved most reliable in estimating the $\gamma d \rightarrow K^{0}_{S}\Lambda(\gamma p)$ events under the proton missing mass distribution.
Momenta of the missing particles was not used to separate the background but is discussed in Section \ref{sec:mcsim}.
This background combined with simulated $K^{0}_{S}\Lambda$ events represented the data fairly well.
Other five track backgrounds that do not produce real $K^{0}_{S}$'s or $\Lambda$'s were significantly reduced by the invariant mass cuts and separated from the signal by a large missing mass.

\begin{comment}
\begin{figure}[htbp]
    \centering
    \begin{subfigure}[t]{0.48\linewidth}
        \includegraphics[width= \linewidth, height = 2.5 in, keepaspectratio = true]{sigma0_150}
        \caption{Shown is a hyperbolic tangent curve fit to the missing mass of $\gamma d \rightarrow K^{0}\Sigma^{0}p \rightarrow\pi^{-}\pi^{+}\pi^{-}p(\gamma p)$ simulation.}
        \label{fig:sigma0_150}
    \end{subfigure}
    ~
    \begin{subfigure}[t]{0.48\linewidth}
        \includegraphics[width= \linewidth, height = 2.5 in, keepaspectratio = true]{signal_150}
        \caption{Shown is a fit to the missing mass of an example data bin. The fit uses a Gaussian and a hyperbolic tangent shape that is parameterized from the simulation.}
        \label{fig:signal_150}
    \end{subfigure}
    \caption{This is an example of a fit to the g13 data to extract the number of events that are missing only one proton.}
    \label{fig:yieldfit}
\end{figure}
\end{comment}

Other hyperon backgrounds were studied using simulations of detector acceptance.
An equal number of events was generated for the $K^{0}_{S}\Lambda$ channel and the 
two lowest energy competing background channels - the $K^{0}_{S}\Sigma^{*}$ 
and $K^*(892)\Lambda$ channels.
Phase space was used for the event generation.
%The final results show a very familiar missing mass distribution.
There was a negligible contribution of both channels, which reflected their extremely low acceptance.
This, combined with the improbability that the missing mass was near the spectator proton mass, suggests that these channels were not contaminating the dataset.

\subsubsection{Four Track Background} \label{sec:fourparticleback}
\begin{figure}[htbp]
    \centering
    \includegraphics[width= \linewidth, height = 3.0 in, keepaspectratio = true]{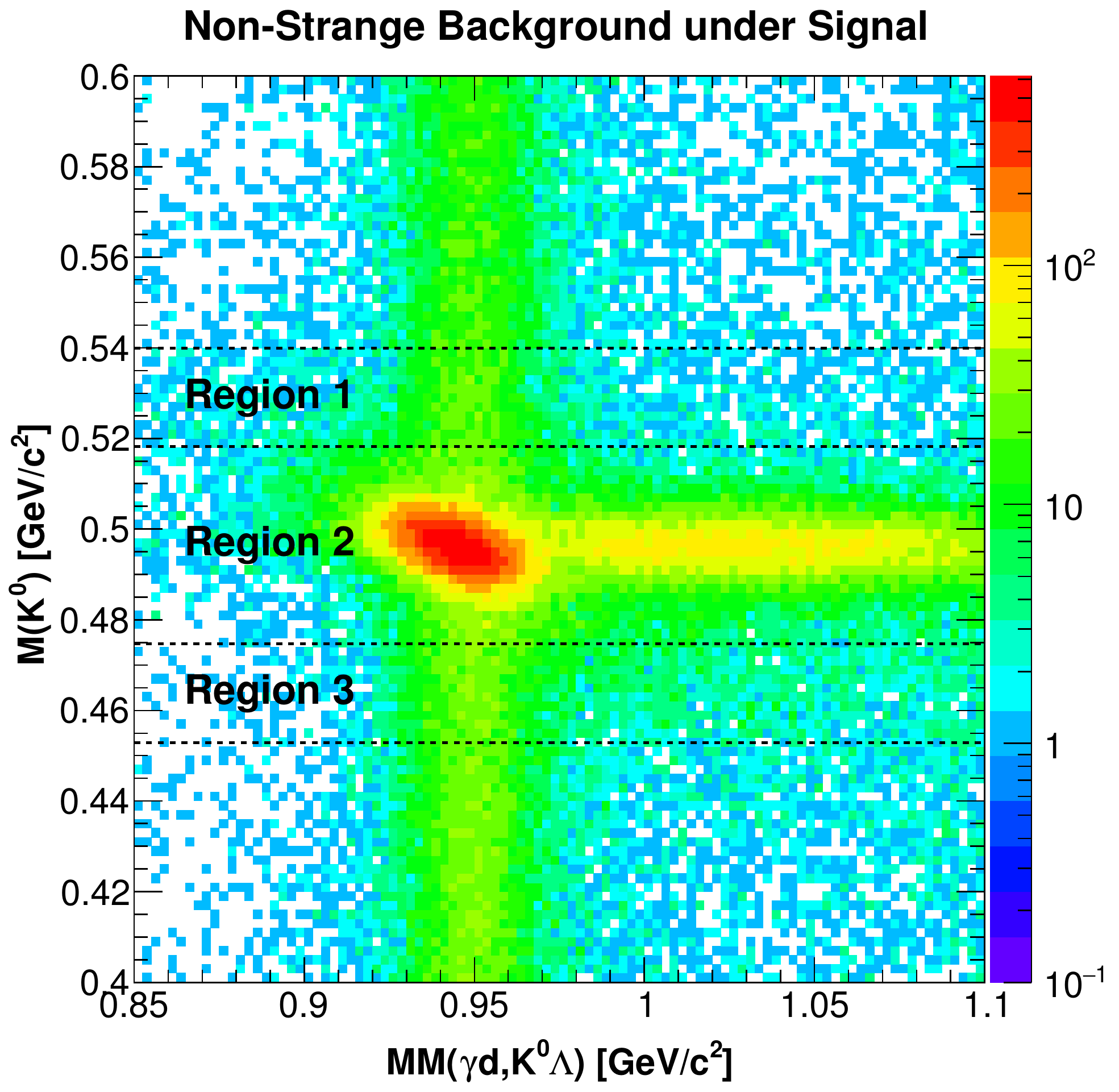}
    
    \includegraphics[width= \linewidth, height = 3.0 in, keepaspectratio = true]{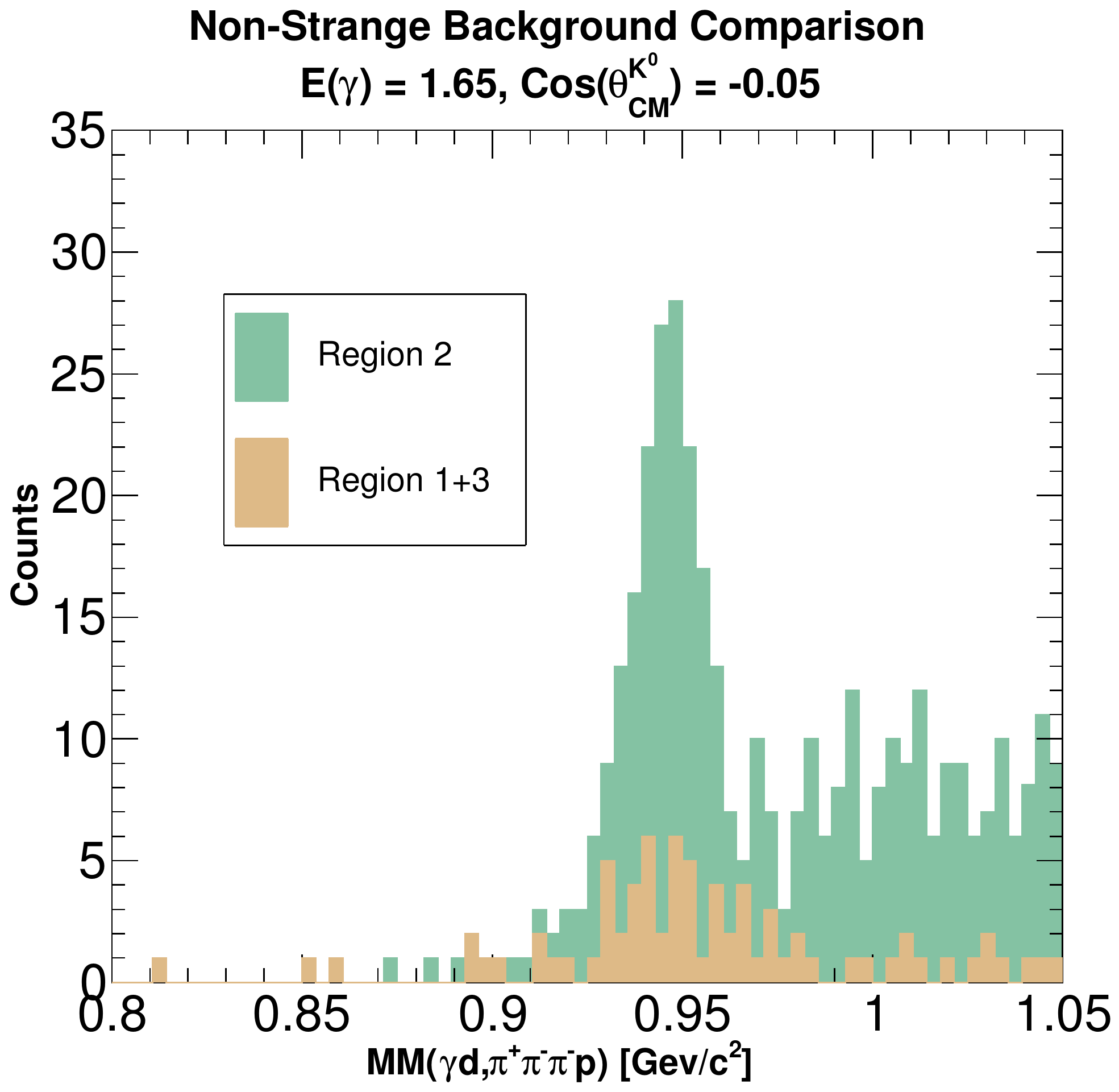}
    \caption{The top panel shows the invariant mass of the 
$\pi^{+}\pi^{-}$ pair versus the $K_S^0 \Lambda$ missing mass. 
This figure demonstrates where non-strange $\pi^{-}\pi^{+}\pi^{-}p$ background 
contributed to the extracted yield. The signal and the $K^{0}_{S}$ sidebands are 
outlined by the dashed lines. 
A small subset of the data (one kinematic bin) is projected onto the $x$-axis in the bottom panel to show a typical background contribution.}
    \label{fig:datasideband}
\end{figure}
While the strange channels (such as $K^{0}_{S}\Lambda$ and $K^{0}_{S}\Sigma^{0}$) were the primary source of our four final state particle events, other processes from non-strange production mechanisms could contribute to the background.
%There are several physical processes that produce this final state.
One to consider is the production channel of $\gamma d \rightarrow \rho\Delta^{0}p\rightarrow \pi^{-}\pi^{+}\pi^{-}p(p)$.
%\subsubsection{$\gamma n \rightarrow \rho\Delta^{0}$}
%The neutral delta resonance can also prove a significant contributor to background.
Both the $\rho$ and $\Delta^{0}$ have a wide mass distribution when compared to either the $K^{0}_{S}$ or $\Lambda$ peaks.
When this channel is considered it could easily produce a relatively broad distribution about the invariant masses of the $K^{0}$ and $\Lambda$.
Likewise if there were other similar background processes, the general trend would be creating a missing mass peak at the value of the proton mass, but would not produce a peak at the kaon or $\Lambda$ mass.
%This channel is shown specifically to give an easy example of a real physics event that could create pion background.

%\subsubsection{$\gamma n \rightarrow \pi^{-}\pi^{+}\pi^{-}p$}

Regardless of the channel, one would expect scattering events where the final state particles were directly produced from photon-nucleon interactions.
In this case, the background from $\gamma d \rightarrow \pi^{-}\pi^{+}\pi^{-}p(p)$ is expected.
Because there were multiple channels contributing to the background, they were modeled with simulations.
This ``random" distribution resulted with kinematics filling in the phase space underneath the signal peaks ($K^{0}_{S}$ and $\Lambda$).
A uniform phase space distribution was generated to model this background.
Although most of the generated phase space events were not in the region of interest, the events that did pass the limiting cuts matched the background shape under the $\Lambda$ signal and the $K^{0}$ signal.

To account for this background, the sidebands of $K^{0}_{S}$ were projected onto the missing mass plane, where by definition this background created a peak at the spectator proton mass.
The number of events that were only missing a spectator proton were found in each region noted in Fig.~\ref{fig:datasideband}.
To obtain the correct number of $K^{0}_{S}\Lambda$ events, subtraction was used based on the sidebands of the $K^{0}_{S}$ distribution.
The events in Region 2 of Fig.~\ref{fig:datasideband} can be written as
\begin{equation}
    N^{K^{0}_{S}\Lambda}_{2} + N^{K^{0}\Sigma^{0}}_{2} + N^{\pi^{-}\pi^{+}\pi^{-}p}_{2},
\end{equation}
where $N^{K^{0}_{S}\Lambda}$ is the number of $\gamma d\rightarrow K^{0}_{S}\Lambda(p)$ events, $N^{K^{0}_{S}\Sigma^{0}}$ is the number of $\gamma d\rightarrow K^{0}_{S}\Sigma^{0}(p)$ events, and $N^{\pi^{-}\pi^{+}\pi^{-}p}$ is the number of $\gamma d\rightarrow \pi^{-}\pi^{+}\pi^{-}p(p)$ events that do not follow from the decays of $K^{0}_{S}$ or $\Lambda$.
To correct for the overestimates of $K^{0}_{S}\Lambda$ yield, $\pi^{-}\pi^{+}\pi^{-}p$ events in Region 1 and Region 3 of Fig.~\ref{fig:datasideband}, 
\begin{equation}
    N^{\pi^{-}\pi^{+}\pi^{-}p}_{1} + N^{\pi^{-}\pi^{+}\pi^{-}p}_{3},
\end{equation}
were subtracted from the events of Region 2.
The size of this background fluctuated near 15\% depending on the kinematic bin. 
This resulted in the raw yield of $K^{0}_{S}\Lambda$ after subtraction of $K^{0}_{S}\Sigma^{0}$ events.

\begin{comment}
\begin{figure}[h]
    \centering
    \includegraphics[width=\linewidth, height = 3in, keepaspectratio = true]{datasideband}
    \caption{Shown is the invariant mass of the $\pi^{+}\pi^{-}$ pair that is marked as a possible kaon versus the missing mass. This figure demonstrates where non-strange $\pi^{-}\pi^{+}\pi^{-}p$ background is contributing to any extracted yield. The signal and the $K^{0}$ sidebands are outlined by the dashed lines.}
    \label{fig:datasideband}
\end{figure}
\end{comment}

\subsection{Photon Flux}
Photons incident on the target were tallied and then corrected by the tagger efficiency as they were written into the flux files \cite{bib:fluxdeter}.
The analysis code then cycled through the files to sort photons into the same energy bin structure as the yield extraction.
%The number of incident photons on the target was measured and recorded in separate data files and required separate processing \cite{bib:fluxdeter}.
Events without a corresponding photon flux file were dropped from the analysis.
%Analysis was also performed on the consistency of the flux files which allowed an estimate of how consistent the run was within itself.
%This was applied into the uncertainty of cross section per bin. 
Analysis was performed on the consistency of the yield-to-flux ratio, or normalized yield.
This generated an estimate of stability for each run within the experiment.
Most energies showed a variation less than 3\% in the normalized yield for g13 
and less than $\approx 5\%$ for g10. 
This uncertainty was accounted for in the calculation of the luminosity uncertainty for the cross section (see Tables \ref{tab:sysuncertg10} and \ref{tab:sysuncertg13}).

\FloatBarrier
\subsection{Monte Carlo Simulation} \label{sec:mcsim}
Monte Carlo simulation was used to determine the true acceptances in the CLAS detector.
In principle, the CLAS detector provides nearly $4\pi$ acceptance, but in reality, the detector has several ``blind'' spots and regions of low efficiency.
Simulation was used to generate $K^{0}_{S}\Lambda(p)$ and $K^{0}_{S}\Sigma^{0} p\rightarrow K^{0}_{S}\Lambda(\gamma p)$ events separately.
Their relative event ratios for each kinematic bin were later weighted in proportions with respect to the real data.
For this study, \textit{fsgen} \cite{bib:fsgen} (a FORTRAN code that uses the PYTHIA framework \cite{bib:pythia})  was used for event generation. %cite this reference
Events were produced from a deuterium target and included the associated Fermi momentum.
The reliability of the simulated events was tested through comparisons of each particle's momentum, including the spectator proton.
The generated events were then passed to the standard CLAS detector simulation, GSIM (a GEANT-3-based simulation code suite for CLAS).
%Once the quality of the generated events had been verified, they were passed to the standard CLAS detector simulation (GSIM)\cite{bib:GSIMGPP}.
The GSIM package uses GEANT to propagate the particles through a simulated CLAS system. %cite this reference
%After event generation, the simulated particles were sent through a model of the detector with GSIM.
It was important to correct for the detector inefficiencies, before the event quantities were sent through the reconstruction algorithm and analyzed.
We used the GPP (GSIM post processor) code
that served two primary purposes: it removed some tracks to 
correct for the inefficiencies in the CLAS detector system at the time of 
the experiment and it smeared the track resolution through the drift chambers 
to better model the position uncertainty of detectors in the 
experimental data.
The trajectories and energies of the final-state particles were recorded 
into the data banks as individual measurements of sub-detector systems.
The files containing the simulation data had the same structure as 
the data files, with the addition of the generated information for each track.

\begin{figure}[h]
    \centering
      \includegraphics[width=\linewidth, height = 2.5in, keepaspectratio = true]{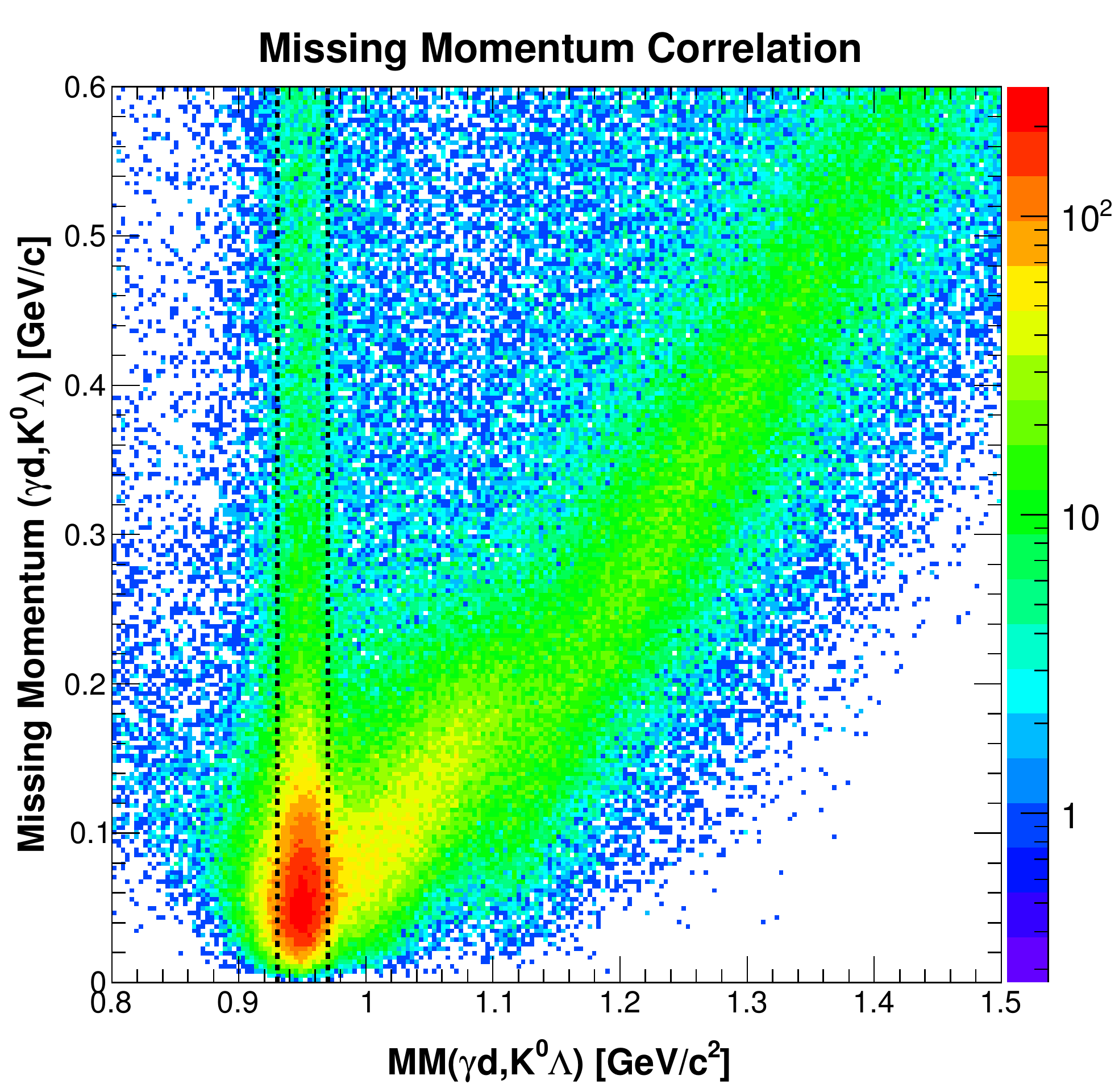}
    \caption{(g13 data) The missing momentum versus the missing mass, $MM(\gamma d,\pi^{+}\pi^{-}\pi^{-}p)$. The missing spectator proton can be seen by the vertical distribution inside the dotted lines, while the diagonal distribution implies events where an extra particle exists within the reaction ($\gamma$ in the case of $\Sigma^{0}$ or $\pi^{0,+}$ in the case of $\Sigma^{*}$/$K^{*}$).}
        \label{fig:6}%
\end{figure}

\begin{figure}[h]
    \centering
    \includegraphics[width=\linewidth,height = 2.5in, keepaspectratio = true]{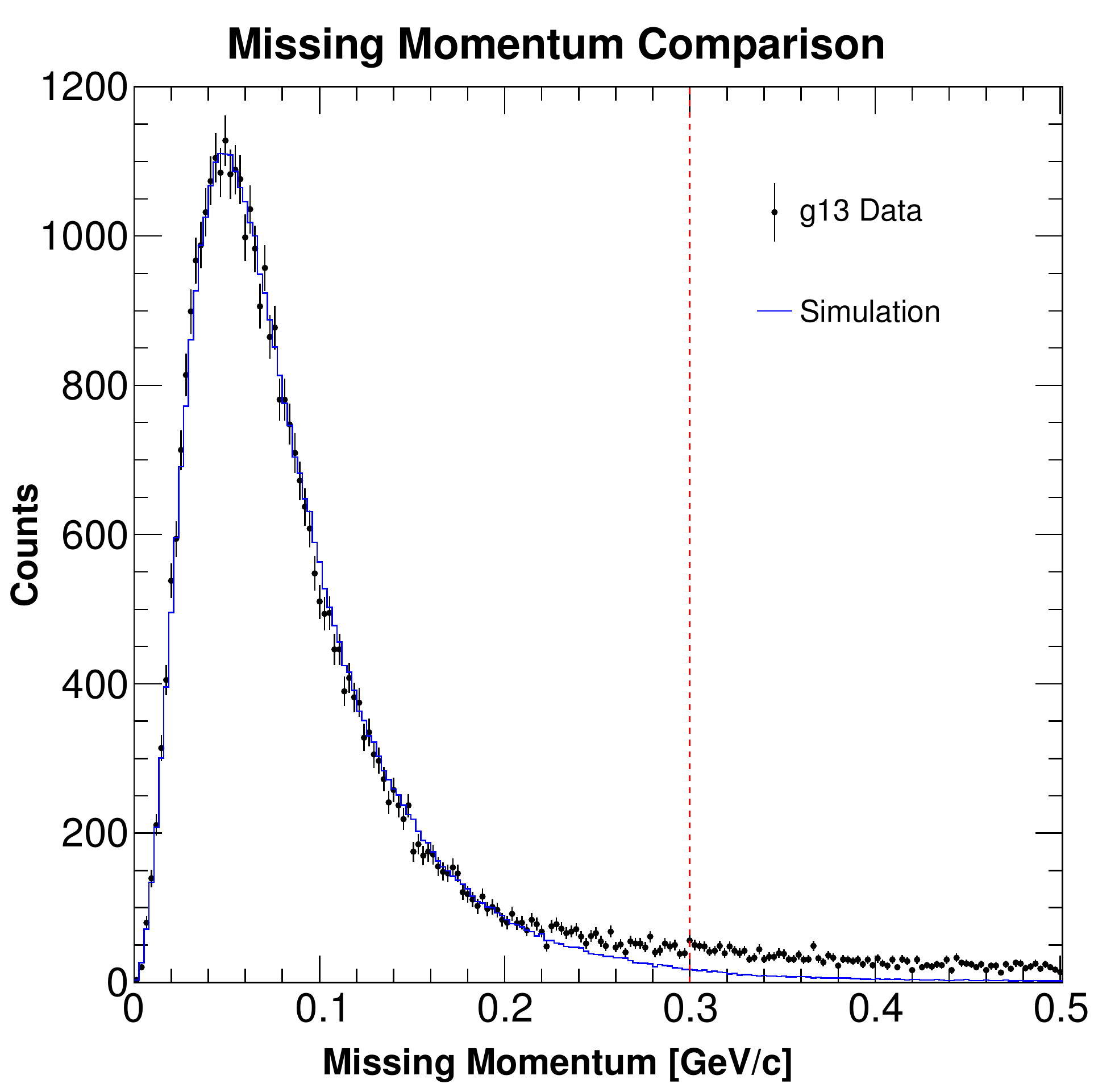}
    \caption{The missing momentum distribution of events in the g13 data (points) and the g13 simulation (line) with a strict cut on missing mass. The simulation was scaled to the data. The vertical red line represents the missing momentum cut applied to reduce final state interaction contributions.}
    \label{fig:missmom}%
\end{figure}

%\subsubsection{Event Generation}
The momentum of the spectator proton was compared to the reconstructed simulation versus data. 
%One of the comparisons between the reconstructed simulation versus data, was that of the momentum of the spectator proton.
The generator began by first selecting the photon energy in the event.
%The event generation began with the selection of photon energy.
With this energy, the Fermi momentum was determined using the Bonn distribution as a weighting factor.
The Bonn potential is based on the exchange of mesons between the nucleons \cite{bib:BonnPotential}. 
The center-of-mass energy, along with all the momenta contained in each generated event, was affected by the Fermi momentum.
The missing momentum in this analysis was described by
\begin{equation}
    |\vec{p}_{p^{spec}}| = |\vec{p}_{\gamma} + \vec{p}_{d} - \vec{p}_{K^{0}_{S}} - \vec{p}_{\Lambda}|,
\end{equation}
where $\vec{p}$ is the momentum vector of each particle: proton, photon, deuteron, kaon, and Lambda.
This missing momentum in each four track event (assuming no missing tracks) represented the Fermi momentum of the undetected spectator proton.
One can see this in the data only if a strict cut on missing mass is applied to remove a significant portion of the $\Sigma^{0}$ background (see Fig.~\ref{fig:6}).
%if we keep this figure it should be improved
%higher quality, tick marks all around
%possibly vertical lines at pm 20 MeV
Applying a cut of $\pm20$~MeV about the expected missing mass peak, $MM(\gamma d,\pi^{+}\pi^{-}\pi^{-}p)$, at the spectator proton results in Fig.~\ref{fig:missmom}.
The agreement between simulation and data confirmed that the weighting of Fermi momentum in event generation appropriately describes the process in quasi-free events.
%add g10 figure here as well

\subsection{Systematic Uncertainties}
Systematic uncertainties were determined for each portion of the experiment.
This includes uncertainties in the target and detector geometries, and 
effects of event selection and cuts.
Most components contributing to the uncertainties were compiled per 
kinematic bin.
The largest uncertainties were associated with forward angles, where a 
blind spot exists from the detector geometry, and at backward angles, 
where statistics and detector efficiencies were poor. 
The average point-to-point uncertainties can be seen in 
Tables \ref{tab:sysuncertg10} and \ref{tab:sysuncertg13}.
These were separated into broad categories to give some sense of the 
source of uncertainty.

The systematic uncertainties underwent extensive internal review, and 
were examined for different choices for analysis cuts 
and different methods of background subtraction for the yield extraction. 
Details are given in Ref. \cite{bib:g10note} and \cite{bib:g13note}.
In addition, one of the largest uncertainties is due to the luminosity. 
This was studied extensively in Ref. \cite{bib:g10ppim} for the g10 
experiment and similar studies were repeated for the g13 experiment \cite{bib:g13note}. 
In general, these systematic uncertainties are typical when compared 
with other CLAS experiments \cite{bib:g11kplam,bib:g1ckplam,bib:MCNabb}. 

\begin{table}[h]
\centering
\caption{Summary of the g10 systematic effects, estimating a total average point-to-point uncertainty \cite{bib:g10note}.}
\label{tab:sysuncertg10}
\begin{tabular}{|c|c|}
\hline
{\bf Investigated Cut} & {\bf Systematic Uncertainty} \\ \hline
Luminosity                 &    5.0\% \\ \hline
Acceptance                 &    1.6\%  \\ \hline
Yield Extraction           &    6.3\%  \\ \hline
Detector                   &    5.0\% \\ \hline % Fiducial and Sector dependence
Branching Ratios           &   $<$1.0\% \\ \hline
{\bf Total}                &   {\bf 10\%}    \\ \hline
\end{tabular}
\end{table}

\begin{table}[h]
\centering
\caption{Summary of the g13 systematic effects, estimating a total average point-to-point uncertainty \cite{bib:g13note}.}
\label{tab:sysuncertg13}
\begin{tabular}{|c|c|}
\hline
{\bf Investigated Cut} & {\bf Systematic Uncertainty} \\ \hline
Luminosity                 &    2.6-7.0\% \\ \hline
Acceptance                 &    1.9-2.1\%  \\ \hline
Yield Extraction           &    4.5-11.4\%  \\ \hline
Detector                   &    3.2\% \\ \hline % Fiducial and Sector dependence
Branching Ratios           &   $<$1.0\% \\ \hline
{\bf Total}                &   {\bf 7\% - 14\%}    \\ \hline
\end{tabular}
\end{table}
\FloatBarrier
%_____________________________EXPERIMENTAL_RESULTS________________________
\section{Experimental Results}
\subsection{Model of the $K^{0}\Lambda$ Differential Cross Section} \label{sec:model}
Several models, such as e.g. KAON-MAID~\cite{bib:kaonmaid}, have been developed for the kaon 
photoproduction channels.
However, while most model calculations for $K^+$ photoproduction off the 
proton show little variation, due to the availability of good quality data,
the $\gamma n \to K^0\Lambda$ predictions from KAON-MAID are largely 
unconstrained.
The combination of the $\gamma n$ and $\gamma K^{0}$ vertices make this channel particularly hard to predict.
The inclusion and exclusion of $t$-channel kaon exchange in calculations 
from KAON-MAID changes the cross section output by large factors (variations 
up to a factor of ten as shown in Ref. \cite{bib:kaonmaid}).
The present data should give enough constraints to tie down several coupling 
strengths that would not only improve other predictions, 
but possibly even allow classification of specific resonances based on 
extracted helicity couplings.

%\onecolumngrid

\begin{figure*}[btph]
    \centering
      \includegraphics[width=\textwidth, keepaspectratio=true]{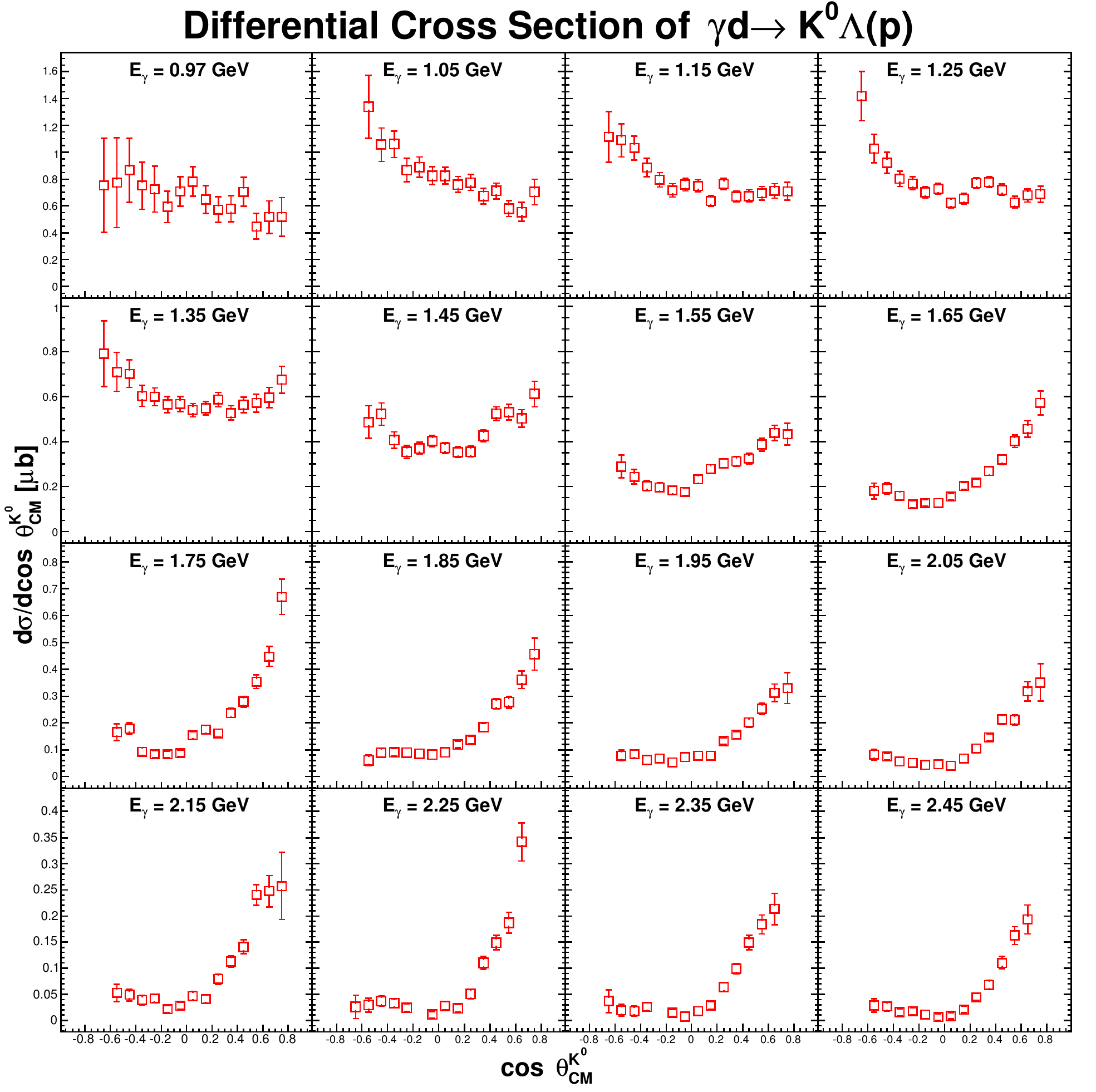} %
    \caption{The g13 differential cross sections in bins of 
$\cos \theta^{K^{0}}_{CM}$ for each beam energy. 
Only the statistical uncertainties are shown.}
        \label{fig:XScosineTheta}%
\end{figure*}

\begin{figure*}[btph]
    \centering
      \includegraphics[width=\textwidth, keepaspectratio=true]{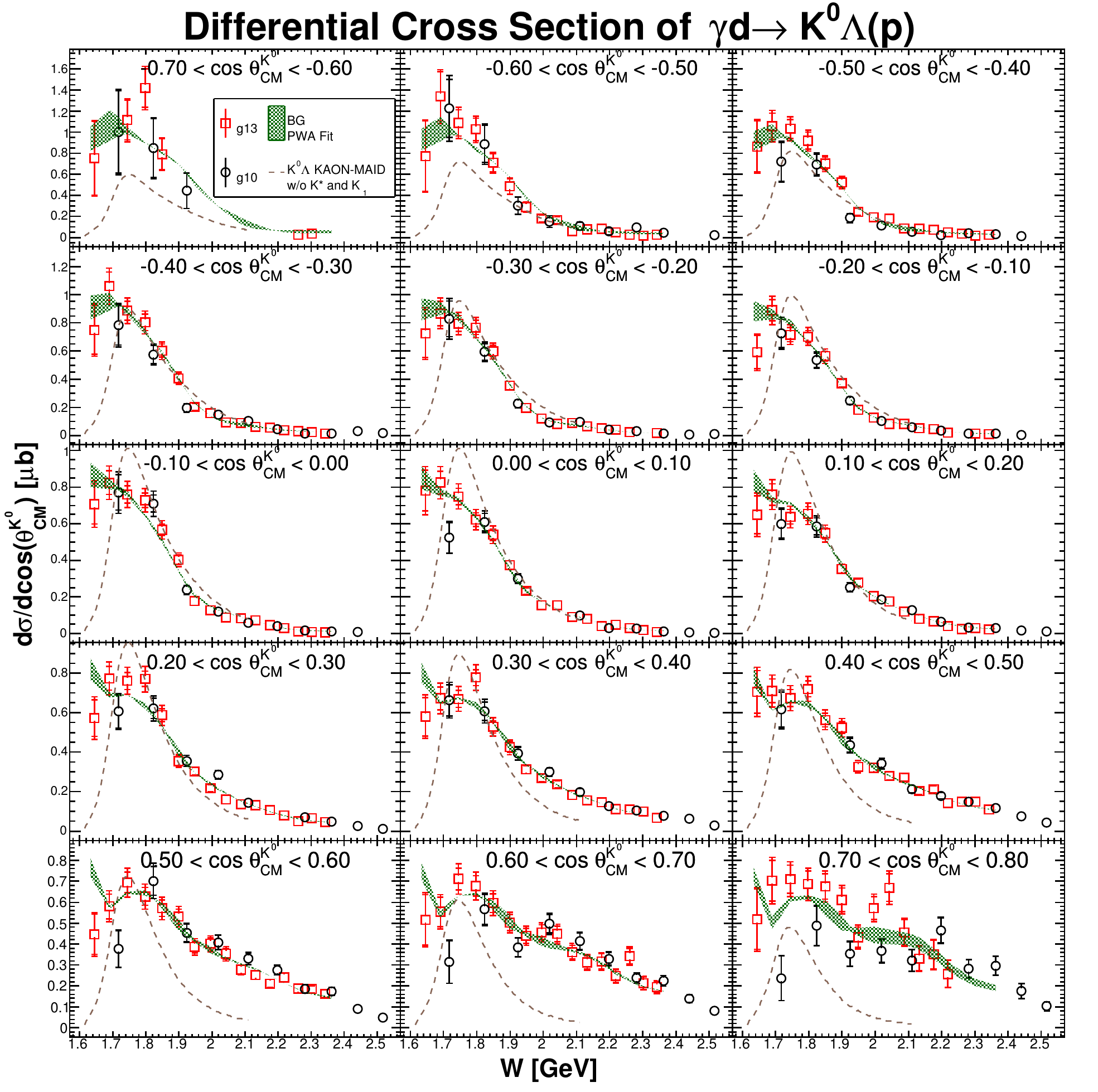}
    \caption{The g10 (circles) and g13 (squares) differential cross sections 
as a function of $W$ (GeV) for each $\cos \theta^{K^{0}}_{CM}$. 
The KAON-MAID model (dashed line) is shown, assuming no contributions from 
the $K^{*}(892)$ and $K_{1}(1270)$, along with the differences between the 
two Bonn-Gatchina $K^{0}\Lambda$ fits (shaded curve). The inner error bars represent 
the statistical uncertainty. The outer error bars are the statistical and 
systematic uncertainties added in quadrature.}
        \label{fig:XSenergyW}%
\end{figure*}
%\twocolumngrid{}

\begin{figure}[tbh]
    \centering
      \includegraphics[height=2in,width=\linewidth, keepaspectratio=true]{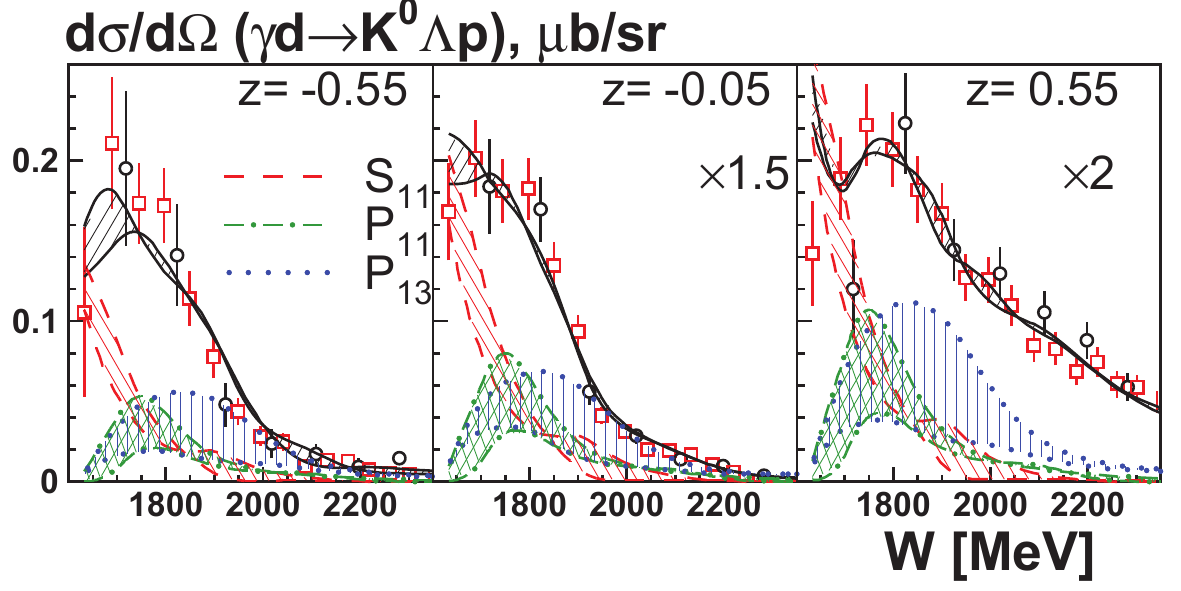}
    \caption{Two PWA solutions from fits using the Bonn-Gatchina model are 
shown to the differential cross section of $\gamma d\to K^{0}\Lambda(p)$ 
at three $z = cos \theta_{CM}^{K^{0}}$ angular bins. The data symbols are 
as given in previous figures. A description of the curves is given by the 
legend and in the text. Parameters of the fits will 
be presented in a future publication.}
        \label{fig:PWAk0lamDCS}%
\end{figure}
\begin{comment}
\begin{figure}[tbh]
    \centering
      \includegraphics[width=\linewidth, keepaspectratio=true]{gn_klam_tot}
    \caption{Shown are two integrated fit solutions (lines) compared to the individually scaled fit solutions at each energy (points) of $\gamma d\rightarrow K^{0}\Lambda(p)$.}
        \label{fig:PWAk0lamTCS}%
\end{figure}
\end{comment}
\subsection{$K^{0}\Lambda$ Differential Cross Section} \label{sec:diffcross}
The luminosity, 
\begin{equation}
    L(E_{\gamma}) = \frac{\Phi (E_{\gamma})\rho \ell N_{A}}{A_{target}}, \label{eq:luminosity}
\end{equation}
where
$E_{\gamma}$ is the beam energy,
$A_{target}$ is the atomic weight of the target,
$\Phi (E_{\gamma})$ is the photon flux,
$\rho$ is the density of the target,
$\ell$ is the length of the target, and
$N_{A}$ is Avogadro's number,
is measured as a function of beam energy in each experiment.
The differential cross section for the $\gamma d \rightarrow K^{0}\Lambda (p)$
reaction can then be written as: 
\begin{equation} 
\frac{d\sigma}{d\cos\theta^{CM}_{K^{0}}} = 
\frac{1}{L(E_{\gamma})\Delta \cos{\theta}_{CM}^{K^{0}}}
\frac{Y\left(E_{\gamma},\cos{\theta}_{CM}^{K^{0}} \right)}{\alpha\left(E_{\gamma},\cos{\theta}_{CM}^{K^{0}}\right)}   
\times B.F.,
%\frac{\Gamma_{K^{0}}}{\Gamma_{K^{0}}\rightarrow K_{S}}
%\frac{\Gamma_{K_{S}}}{\Gamma_{K_{S}}\rightarrow\pi^+\pi^-}\frac{\Gamma_{\Lambda }}{\Gamma_{\Lambda}\rightarrow p\pi^-},
\label{eq:diffcross}                                                
\end{equation} 
where
$\Delta\cos({\theta}_{CM}^{K^{0}})$ is the bin width of $\cos{\theta}_{CM}^{K^{0}}$,
$Y\left(E_{\gamma},\cos{\theta}_{CM}^{K^{0}} \right)$ is the corrected yield,
$\alpha\left(E_{\gamma},\cos{\theta}_{CM}^{K^{0}}\right)$ is the CLAS acceptance,
and $B.F.$ is the branching fraction or inverse branching ratios of the decay channels for the neutral hadrons ($K^{0}\rightarrow K^{0}_{S}$, $K^{0}_{S} \rightarrow \pi^{+}\pi^{-}$, and $\Lambda \rightarrow \pi^{-}p$).
%Ideally the differential cross section would be measured as a function of $W$, rather than $E_{\gamma}$.
%Although $W$ could be found for each event, the photon flux is independent of the Fermi momentum inside the deuteron.
%The final data points are positioned on the event median $E_{\gamma}$.
%In most cases $E_{\gamma}^{median}$ corresponded to the geometric mean of $E_{\gamma}$ in each cross section bin.
Using the g13 data set, Fig.~\ref{fig:XScosineTheta} shows the differential cross section of the $\gamma d \rightarrow K^{0}\Lambda (p)$ reaction with respect to $\cos \theta^{K^{0}}_{CM}$ for 100~MeV photon energy bins between 0.9~GeV to 2.5~GeV.

Preliminary fits using PWA from the Bonn-Gatchina group were 
applied to the measured data \cite{bib:BoGa}.
The $s$-channel diagrams, where $N^*$ resonances form, contain two main variables.
The first unknown is that of the resonance decay, $N^{*}\to K^{0}\Lambda$.
This can be restricted by utilizing previous $K^+ \Lambda$ fit results from 
proton targets.
The second and more interesting unknown is that of $\gamma n \to N^{*}$.
Not only is $\gamma n \to N^{*}$ different from that of $\gamma p \to N^{*}$, 
due to the photocouplings, 
but not all resonances will have a strong decay to the $KY$ channels.
To best describe the underlying processes, this PWA employed a multi-channel 
fit that incorporated observables from 
$\gamma d\to \pi^{-}p(p)$, $\pi^{-}p\to \gamma n$, 
$\gamma d\to \pi^{0}n(p)$, $\gamma d\to \eta n(p)$, and 
$\gamma d\to K^{+}\Sigma^{-}(p)$.
As a result of the preliminary fit, two main solutions were found to 
describe the data.
Both solutions seem to describe $\gamma d\to K^{+}\Sigma^{-}(p)$ and 
$\gamma d\to K^{0}\Lambda(p)$ reasonably well, as shown below. 

In Fig.~\ref{fig:XSenergyW},  the cross section is shown as a function 
of center-of-mass energy for various $\cos \theta^{K^{0}}_{CM}$ bins 
including both the g10 and g13 data sets.
Close agreement is seen between the two experiments, with some discrepancies 
in the forward bin: 0.7 $< \cos \theta^{K^{0}}_{CM} <$ 0.8.
Although the exact cause of the small difference in this forward bin is 
unknown, it is assumed that this demonstrates the uncertainty of modeling 
the detector and field map in this kinematic regime 
(two of the main differences between these experiments were the magnitude 
and directionality of the magnetic field).
The KAON-MAID model is also shown, assuming no contributions from the 
$K^{*}(892)$ and $K_{1}(1270)$.
These parameters were chosen for the model as this provided the best 
agreement with data.
From this it is seen that these data will be essential to better constrain 
$t$-channel contributions.
The complementary nature of $\gamma d \rightarrow K^{0}\Lambda (p)$ compared 
to $\gamma p \rightarrow K^{+}\Lambda$, where one has a neutral exchange 
in the $t$-channel and the other a charged exchange, can help 
differentiate between contributions from various $t$-channel exchanges
(and the interference between $s$-channel and $t$-channel diagrams).

The  cross sections of the data are in good agreement with 
the PWA fits done by the Bonn-Gatchina group \cite{bib:BoGa} as
shown in Figs.~\ref{fig:XSenergyW} and \ref{fig:PWAk0lamDCS}.
In the latter, the shaded regions show the range of contribution 
from different $s$-channel partial waves ($S_{11}$, $P_{11}$ and $P_{13}$ 
denoted in the legend of the figure) that contribute to the total strength
(shown by the solid lines). 
Further work on measurements of photoproduction observables off the 
deuteron will help differentiate between the two Bonn-Gatchina solutions 
shown here. Such work is in progress and will be presented in a 
separate publication.

\FloatBarrier
\subsection{Total Cross Section}
The total cross section can be found by integrating over all 
$\cos \theta^{K^{0}}_{CM}$ of the differential cross section.
This has two sources of uncertainty: that of the fit to the 
data points, and that associated with the absence of data 
at extreme angles.

\begin{figure}[tbh]
    \centering
      \includegraphics[width=\linewidth,keepaspectratio = true]{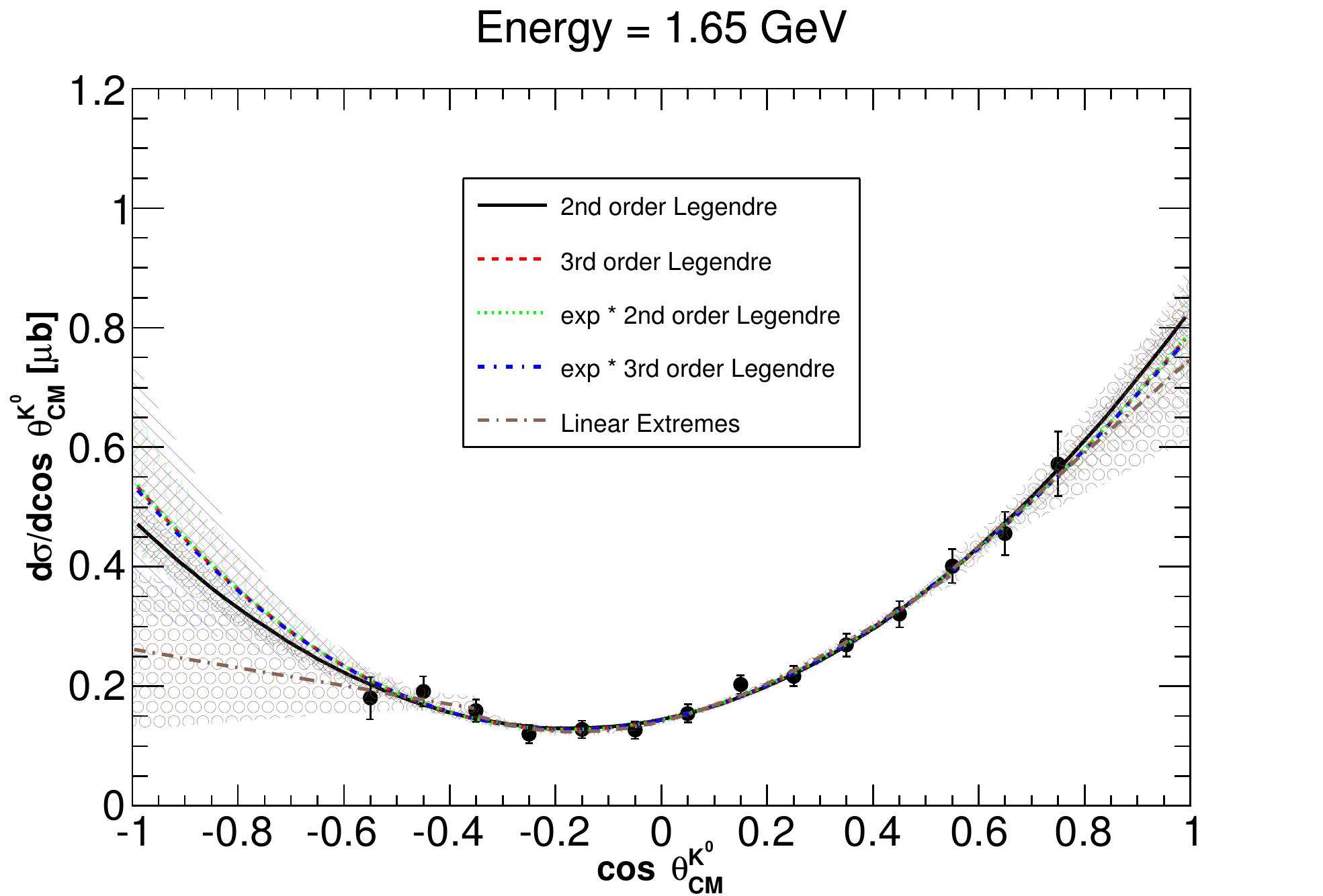}
    \caption{Different fits of the g13 differential cross sections 
(at $E_{\gamma} = 1.65$~GeV) using five different fits. 
The shaded regions represent a 1$\sigma$ band of the fit to the data.}
        \label{fig:CrossfitperE}%
\end{figure}

\begin{figure}[tbh]
    \centering
      \includegraphics[width=\linewidth,keepaspectratio = true]{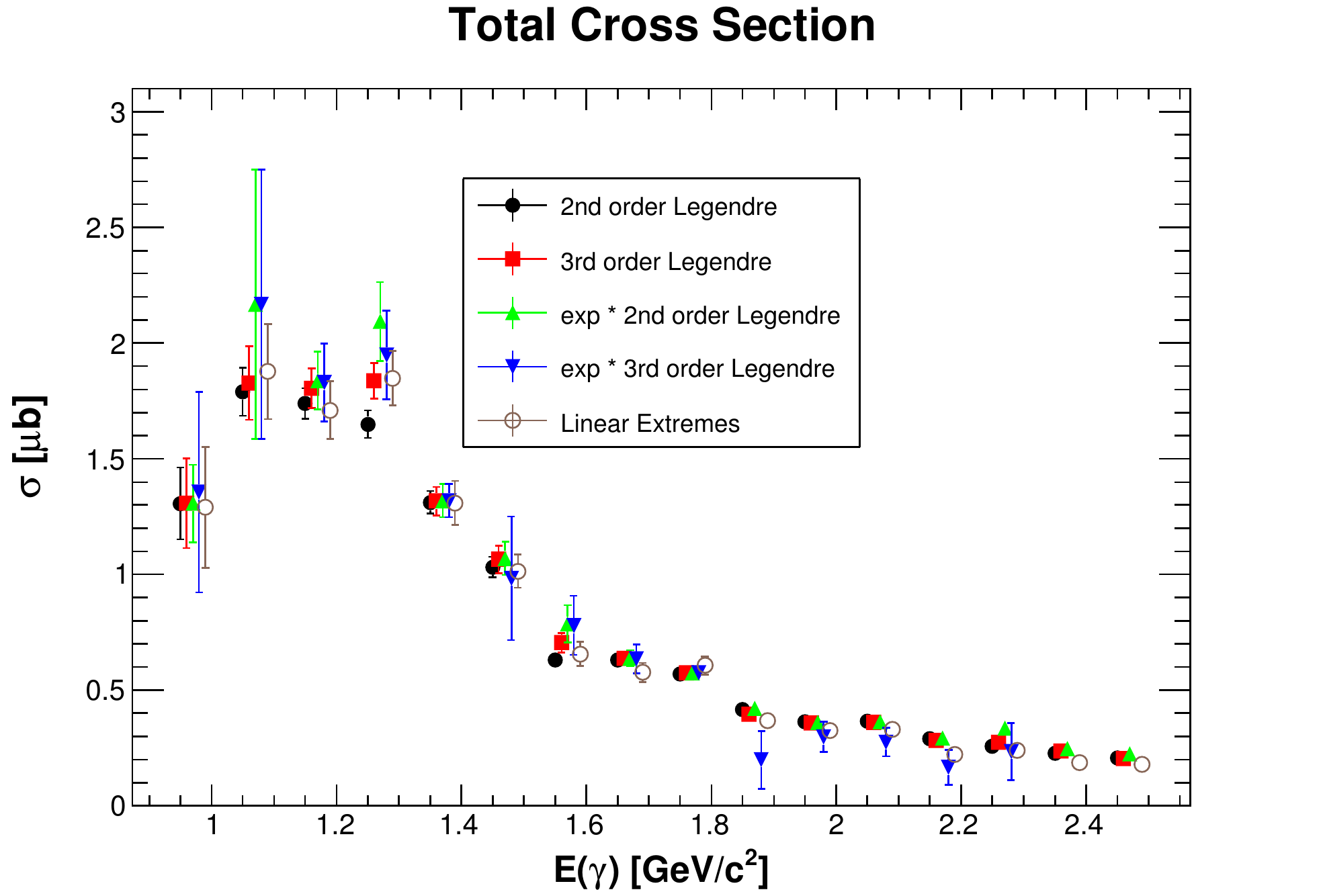}
    \caption{The integration of each of the different fit functions
used for the g13 differential cross sections.}
        \label{fig:totcrossfit}%
\end{figure}

\begin{figure*}[bth]
    \centering
    \includegraphics[width=0.49\textwidth,keepaspectratio = true]{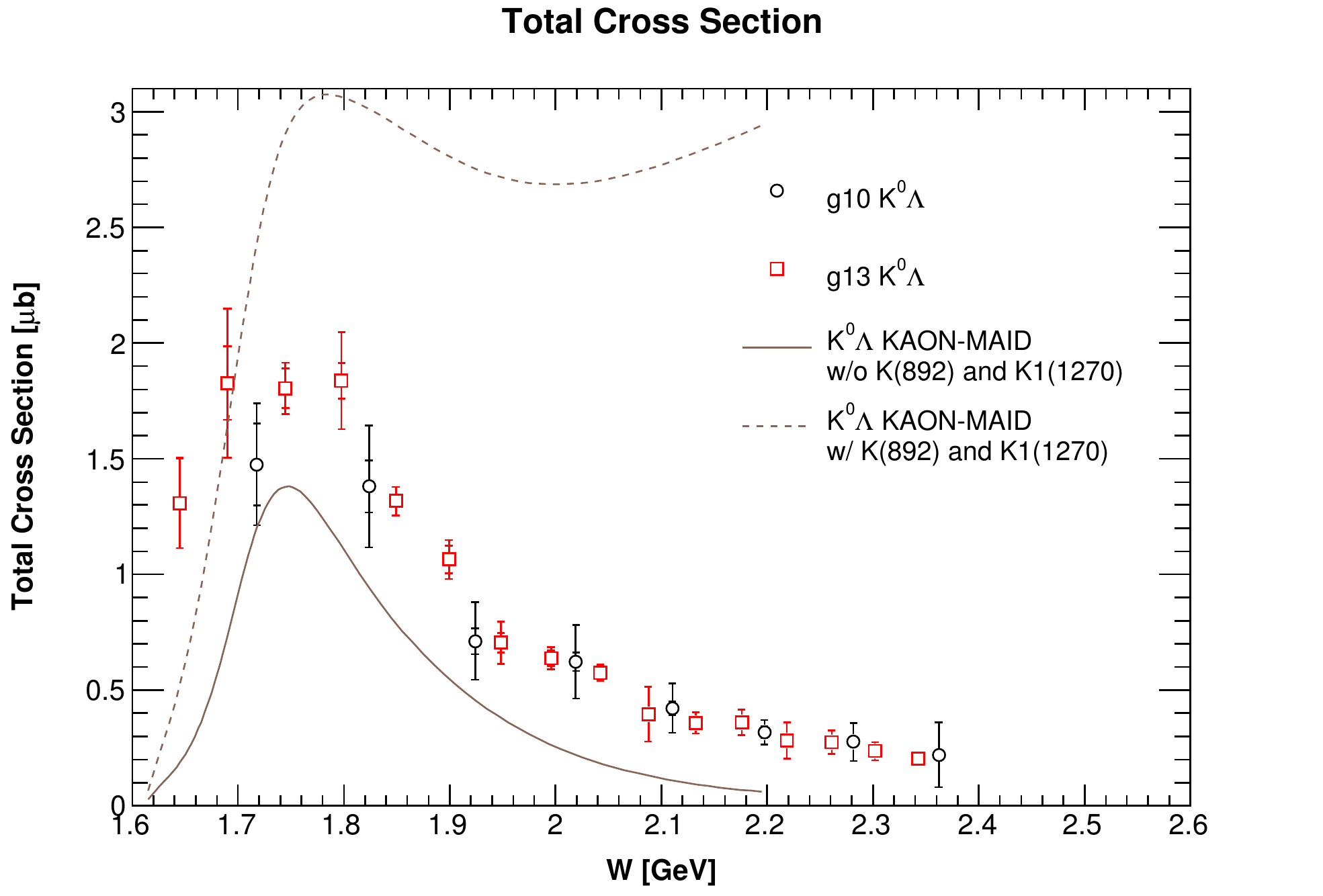}
    \includegraphics[width=0.49\textwidth,keepaspectratio = true]{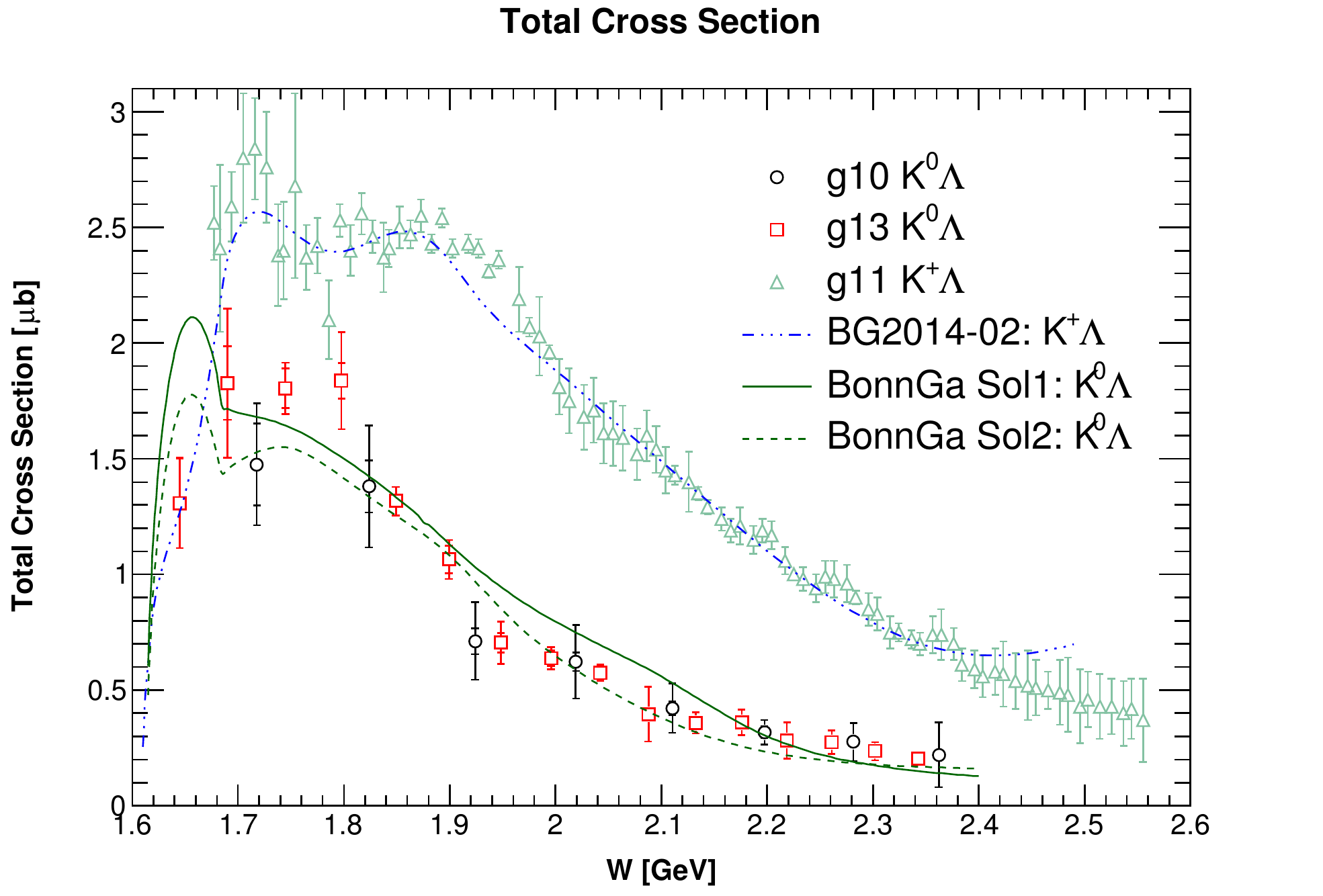}
    \caption{Shown in the left panel are comparisons of the data to the KAON-MAID 
predictions, with two different input parameters (dashed and solid lines). 
This prediction clearly is not very constrained in this channel, in reference 
to contributions from higher mass kaons. The right panel shows the total 
cross section of $\gamma p \rightarrow K^{+}\Lambda$ from the CLAS g11 dataset~\cite{bib:g11kplam}
compared to $\gamma d \to K^{0}\Lambda(p)$.  The total cross section results from both the
g13 (squares) and g10 (circles) experiments.}
    \label{fig:totcrossWb}%
\end{figure*}

Despite the fact that an individual fit function may fit the data within 
the measured angular region quite well, it may not be fully representative 
of the overall uncertainty.
To obtain an estimate on the uncertainty attributed with extrapolations to 
extreme $\cos \theta^{K^{0}}_{CM}$ regions, many functions were tried.
These functions included:
\begin{itemize}
    \itemsep0em 
    \item A second order Legendre polynomial 
    \item A third order Legendre polynomial 
    \item A second order Legendre polynomial multiplied by an exponential 
    \item A third order Legendre polynomial multiplied by an exponential 
    \item A third order Legendre polynomial with linear extrapolations 
\end{itemize}
These functions can fit the data well and be assumed to span a variation of 
realistic behaviors near the forward and backward angles.
The uncertainty of the integration incorporated the covariance matrix given 
by the fit.
The larger the error bands in the range of $\cos\theta^{K^{0}}_{CM}$ 
from $-1$ to 1, the larger the uncertainty in the integration.
Fig.~\ref{fig:CrossfitperE} demonstrates several fits to the data with a 
1$\sigma$ error band.
The integrated cross sections for each fit can be seen in 
Fig.~\ref{fig:totcrossfit}.
The quoted total cross section uses the third order Legendre polynomial.
The base fit is shown in Fig.~\ref{fig:totcrossWb}.
The inner error bars are the uncertainty estimates from the third order 
Legendre polynomial integration.
The outer error bars represent the computed standard deviation (between 
third order polynomial and all other fits) added in quadrature with the 
inner error bars.

Previous analyses of $\gamma p \rightarrow K^{+}\Lambda$ \cite{bib:BoGa} 
have shown that there is at least one $s$-channel resonance necessary to 
describe the data that was not needed for PWA of the pion data.
Therefore, the channel $K^{0}\Lambda$ should be able to confirm these 
found states.
For example, Fig.~\ref{fig:totcrossWb} clearly shows a ``bump" in the 
$\gamma p \to K^{+}\Lambda$ channel near 1900~MeV often attributed to 
$N(1900)3/2^+$.
This enhancement is not seen in $\gamma d\to K^{0}\Lambda(p)$, albeit with 
fewer data points available.
This suggests that this effect is due to missing interference terms. 
One interpretation is to view these missing terms as contributing to the 
excess $K^{+}\Lambda$ cross section through the interference of a resonant 
state, the $N(1900)3/2^+$ and $t$-channel background processes.
This is assumed since $\gamma d\rightarrow K^{0}\Lambda(p)$ has a 
suppression of $t$-channel terms \cite{bib:backwardpeak}, described by 
kaon exchange, which should make this reaction ideal for identifying $N^*$ resonances.
This implies that partial wave analyses combined with the nature of 
$\gamma d\to K^{0}\Lambda(p)$ production will be able to provide constraints 
for models describing nucleon resonances that couple strongly to the $KY$ 
decay channels.

%_____________________________CONCLUSIONS________________________
\section{Conclusions}
In summary, the differential and total cross sections of 
$\gamma d\to K^{0}\Lambda(p)$ have been presented from two different 
CLAS experiments, which are in good agreement.
Due to the fact that previous data on this channel are scarce, the majority 
of presented kinematics are the first of their kind.
These data have allowed a preliminary PWA fit to be completed, which produced 
two independent solutions to describe the intermediate processes.
The PWA are being extended to fit both the present results and the previous 
$K^{+}\Lambda$ results and other available data, with the goal of 
investigating whether existing $s$-channel $N^*$ resonances can provide a 
reasonable description of these data, and perhaps to further constrain 
the pole properties of these $N^*$'s.

These data contain unique information that can be extracted to help with 
resonance classification and determination of helicity amplitudes,
for example, in the contributions of the $N(1900)3/2^+$ resonance in 
strangeness photoproduction.
Clearly more investigation is needed to correctly describe the nucleon 
excitation spectrum.
It is expected that the continued study of observables in this channel will 
be able to identify the best PWA solutions that can fit the data.
The identification of the correct fit will improve our current understanding 
of the $s$-channel contributions to $KY$ cross sections.

\begin{acknowledgments}
The authors gratefully acknowledge the work of Jefferson
Lab staff in the Accelerator and Physics Divisions. This
work was supported by: the United Kingdom’s Science
and Technology Facilities Council (STFC); the
Chilean Comisi\`on Nacional de Investigaci\`on Cient\`ifica y Tecnoĺ\`ogica (CONICYT); the Italian Istituto Nazionale di Fisica Nucleare; the French Centre National de la Recherche
Scientifique; the French Commissariat \`a l’Energie Atomique;
the U.S. National Science Foundation; and the National
Research Foundation of Korea. Jefferson Science Associates,
LLC, operates the Thomas Jefferson National Accelerator
Facility for the the U.S. Department of Energy under Contract
No. DE-AC05-06OR23177. %We also thank Eberhard Klempt, Andrei Sarantsev, and Victor Nikonov
%for providing calculations from the refit Bonn-Gatchina partial wave analysis.
\end{acknowledgments}

%________________________________________________________
%appendix on Cross section numbers
\FloatBarrier
\appendix

%\onecolumngrid

\section{Table of Cross Sections}
%\twocolumngrid

\begin{longtable}{|ccccc|}
\caption{G10 differential cross section.}\\
\toprule
{\bf \boldmath$E_{\gamma}$} &
{\bf \boldmath$\cos\theta^{K^{0}}_{CM}$ } &
%{\bf \boldmath$\frac{d\sigma}{d\cos\theta}$} &
{\bf \nicefrac{\boldmath$d\sigma$}{\boldmath$d\cos\theta$}} &
{\bf \boldmath$\delta$(Stat.)} &
{\bf \boldmath$\delta$(Syst.)}\\
{\bf (GeV)} &
{\bf  } &
{\bf (\boldmath$\mu b$)} &
{\bf (\boldmath$\mu b$)} &
{\bf (\boldmath$\mu b$)}\\
\midrule\endfirsthead
\caption[]{(continued) G10 differential cross section.}\\
\toprule
{\bf \boldmath$E_{\gamma}$} &
{\bf \boldmath$\cos\theta^{K^{0}}_{CM}$ } &
%{\bf \boldmath$\frac{d\sigma}{d\cos\theta}$} &
{\bf \nicefrac{\boldmath$d\sigma$}{\boldmath$d\cos\theta$}} &
{\bf \boldmath$\delta$(Stat.)} &
{\bf \boldmath$\delta$(Syst.)}\\
{\bf (GeV)} &
{\bf  } &
{\bf (\boldmath$\mu b$)} &
{\bf (\boldmath$\mu b$)} &
{\bf (\boldmath$\mu b$)}\\ \midrule\endhead
1.10 & -0.65 & 1.00013 & 0.39197 & 0.10001 \\ 
1.10 & -0.55 & 1.22589 & 0.27443 & 0.12259 \\ 
1.10 & -0.45 & 0.71969 & 0.18661 & 0.07197 \\ 
1.10 & -0.35 & 0.78350 & 0.14200 & 0.07835 \\ 
1.10 & -0.25 & 0.82721 & 0.12747 & 0.08272 \\ 
1.10 & -0.15 & 0.72508 & 0.10020 & 0.07251 \\ 
1.10 & -0.05 & 0.77049 & 0.09670 & 0.07705 \\ 
1.10 & 0.05 & 0.52375 & 0.08326 & 0.05238 \\ 
1.10 & 0.15 & 0.59885 & 0.07771 & 0.05989 \\ 
1.10 & 0.25 & 0.60626 & 0.08459 & 0.06063 \\ 
1.10 & 0.35 & 0.66305 & 0.07958 & 0.06631 \\ 
1.10 & 0.45 & 0.61605 & 0.09009 & 0.06161 \\ 
1.10 & 0.55 & 0.37751 & 0.08750 & 0.03775 \\ 
1.10 & 0.65 & 0.31476 & 0.10235 & 0.03148 \\ 
1.10 & 0.75 & 0.23623 & 0.10792 & 0.02362 \\ 
\hline
1.30 & -0.65 & 0.84988 & 0.28049 & 0.08499 \\ 
1.30 & -0.55 & 0.88767 & 0.17588 & 0.08877 \\ 
1.30 & -0.45 & 0.69275 & 0.09751 & 0.06927 \\ 
1.30 & -0.35 & 0.57456 & 0.06604 & 0.05746 \\ 
1.30 & -0.25 & 0.59504 & 0.05872 & 0.05950 \\ 
1.30 & -0.15 & 0.53502 & 0.05056 & 0.05350 \\ 
1.30 & -0.05 & 0.71023 & 0.04659 & 0.07102 \\ 
1.30 & 0.05 & 0.60923 & 0.04677 & 0.06092 \\ 
1.30 & 0.15 & 0.58598 & 0.04683 & 0.05860 \\ 
1.30 & 0.25 & 0.62218 & 0.05072 & 0.06222 \\ 
1.30 & 0.35 & 0.60768 & 0.04693 & 0.06077 \\ 
1.30 & 0.55 & 0.70114 & 0.06825 & 0.07011 \\ 
1.30 & 0.65 & 0.56572 & 0.07038 & 0.05657 \\ 
1.30 & 0.75 & 0.48735 & 0.09389 & 0.04874 \\ 
\hline
1.50 & -0.65 & 0.44312 & 0.16679 & 0.04431 \\ 
1.50 & -0.55 & 0.30142 & 0.08267 & 0.03014 \\ 
1.50 & -0.45 & 0.18577 & 0.04228 & 0.01858 \\ 
1.50 & -0.35 & 0.19701 & 0.03086 & 0.01970 \\ 
1.50 & -0.25 & 0.22659 & 0.02968 & 0.02266 \\ 
1.50 & -0.15 & 0.24699 & 0.02800 & 0.02470 \\ 
1.50 & -0.05 & 0.23696 & 0.02521 & 0.02370 \\ 
1.50 & 0.05 & 0.29974 & 0.02627 & 0.02997 \\ 
1.50 & 0.15 & 0.25251 & 0.02437 & 0.02525 \\ 
1.50 & 0.25 & 0.35535 & 0.02824 & 0.03554 \\ 
1.50 & 0.35 & 0.39293 & 0.03049 & 0.03929 \\ 
1.50 & 0.45 & 0.43659 & 0.03435 & 0.04366 \\ 
1.50 & 0.55 & 0.45310 & 0.04106 & 0.04531 \\ 
1.50 & 0.65 & 0.38425 & 0.04451 & 0.03843 \\ 
1.50 & 0.75 & 0.35311 & 0.05939 & 0.03531 \\ 
\hline
1.70 & -0.55 & 0.15057 & 0.05422 & 0.01506 \\ 
1.70 & -0.45 & 0.11339 & 0.02715 & 0.01134 \\ 
1.70 & -0.35 & 0.14933 & 0.02108 & 0.01493 \\ 
1.70 & -0.25 & 0.09314 & 0.01801 & 0.00931 \\ 
1.70 & -0.15 & 0.10314 & 0.01543 & 0.01031 \\ 
1.70 & -0.05 & 0.12015 & 0.01516 & 0.01202 \\ 
1.70 & 0.15 & 0.18413 & 0.01695 & 0.01841 \\ 
1.70 & 0.25 & 0.28597 & 0.02109 & 0.02860 \\ 
1.70 & 0.35 & 0.29966 & 0.02143 & 0.02997 \\ 
1.70 & 0.45 & 0.34241 & 0.02653 & 0.03424 \\ 
1.70 & 0.55 & 0.40805 & 0.03285 & 0.04081 \\ 
1.70 & 0.65 & 0.49763 & 0.04392 & 0.04976 \\ 
1.70 & 0.75 & 0.36749 & 0.05521 & 0.03675 \\ 
\hline
1.90 & -0.55 & 0.11043 & 0.03719 & 0.01104 \\ 
1.90 & -0.45 & 0.05705 & 0.01839 & 0.00571 \\ 
1.90 & -0.35 & 0.10543 & 0.01779 & 0.01054 \\ 
1.90 & -0.25 & 0.09800 & 0.01350 & 0.00980 \\ 
1.90 & -0.15 & 0.05759 & 0.01111 & 0.00576 \\ 
1.90 & -0.05 & 0.05927 & 0.01208 & 0.00593 \\ 
1.90 & 0.05 & 0.09919 & 0.01287 & 0.00992 \\ 
1.90 & 0.15 & 0.12653 & 0.01322 & 0.01265 \\ 
1.90 & 0.25 & 0.14384 & 0.01510 & 0.01438 \\ 
1.90 & 0.35 & 0.19756 & 0.01787 & 0.01976 \\ 
1.90 & 0.45 & 0.21230 & 0.01875 & 0.02123 \\ 
1.90 & 0.55 & 0.33042 & 0.02857 & 0.03304 \\ 
1.90 & 0.65 & 0.41433 & 0.03976 & 0.04143 \\ 
1.90 & 0.75 & 0.32115 & 0.05167 & 0.03211 \\ 
\hline
2.10 & -0.55 & 0.05985 & 0.03258 & 0.00598 \\ 
2.10 & -0.45 & 0.02499 & 0.01678 & 0.00250 \\ 
2.10 & -0.35 & 0.04487 & 0.01100 & 0.00449 \\ 
2.10 & -0.25 & 0.04053 & 0.00926 & 0.00405 \\ 
2.10 & -0.15 & 0.03574 & 0.00842 & 0.00357 \\ 
2.10 & -0.05 & 0.04052 & 0.00825 & 0.00405 \\ 
2.10 & 0.05 & 0.02846 & 0.00848 & 0.00285 \\ 
2.10 & 0.15 & 0.06472 & 0.01016 & 0.00647 \\ 
2.10 & 0.35 & 0.12534 & 0.01356 & 0.01253 \\ 
2.10 & 0.45 & 0.17897 & 0.01659 & 0.01790 \\ 
2.10 & 0.55 & 0.27612 & 0.02319 & 0.02761 \\ 
2.10 & 0.65 & 0.32914 & 0.03134 & 0.03291 \\ 
2.10 & 0.75 & 0.46576 & 0.05962 & 0.04658 \\ 
\hline
2.30 & -0.55 & 0.09393 & 0.02631 & 0.00939 \\ 
2.30 & -0.45 & 0.03974 & 0.01434 & 0.00397 \\ 
2.30 & -0.35 & 0.01601 & 0.00850 & 0.00160 \\ 
2.30 & -0.25 & 0.03098 & 0.00773 & 0.00310 \\ 
2.30 & -0.15 & 0.01400 & 0.00632 & 0.00140 \\ 
2.30 & -0.05 & 0.01652 & 0.00612 & 0.00165 \\ 
2.30 & 0.05 & 0.02725 & 0.00635 & 0.00272 \\ 
2.30 & 0.15 & 0.03218 & 0.00675 & 0.00322 \\ 
2.30 & 0.25 & 0.06879 & 0.00908 & 0.00688 \\ 
2.30 & 0.35 & 0.10418 & 0.01186 & 0.01042 \\ 
2.30 & 0.45 & 0.14920 & 0.01497 & 0.01492 \\ 
2.30 & 0.55 & 0.18561 & 0.01928 & 0.01856 \\ 
2.30 & 0.65 & 0.23703 & 0.02506 & 0.02370 \\ 
2.30 & 0.75 & 0.28270 & 0.04309 & 0.02827 \\ 
\hline
2.50 & -0.55 & 0.04511 & 0.02009 & 0.00451 \\ 
2.50 & -0.45 & 0.03033 & 0.00999 & 0.00303 \\ 
2.50 & -0.35 & 0.01442 & 0.00675 & 0.00144 \\ 
2.50 & -0.25 & 0.01399 & 0.00549 & 0.00140 \\ 
2.50 & -0.15 & 0.01566 & 0.00544 & 0.00157 \\ 
2.50 & -0.05 & 0.01149 & 0.00374 & 0.00115 \\ 
2.50 & 0.05 & 0.00996 & 0.00572 & 0.00100 \\ 
2.50 & 0.15 & 0.02832 & 0.00729 & 0.00283 \\ 
2.50 & 0.25 & 0.04826 & 0.00821 & 0.00483 \\ 
2.50 & 0.35 & 0.07632 & 0.01049 & 0.00763 \\ 
2.50 & 0.45 & 0.11663 & 0.01322 & 0.01166 \\ 
2.50 & 0.55 & 0.17365 & 0.01968 & 0.01736 \\ 
2.50 & 0.65 & 0.22403 & 0.02580 & 0.02240 \\ 
2.50 & 0.75 & 0.29691 & 0.04540 & 0.02969 \\ 
\hline
2.70 & -0.45 & 0.01367 & 0.01099 & 0.00137 \\ 
2.70 & -0.35 & 0.03146 & 0.00891 & 0.00315 \\ 
2.70 & -0.25 & 0.00871 & 0.00492 & 0.00087 \\ 
2.70 & -0.15 & 0.00528 & 0.00365 & 0.00053 \\ 
2.70 & -0.05 & 0.00695 & 0.00407 & 0.00070 \\ 
2.70 & 0.05 & 0.00555 & 0.00455 & 0.00056 \\ 
2.70 & 0.15 & 0.01532 & 0.00523 & 0.00153 \\ 
2.70 & 0.25 & 0.02582 & 0.00634 & 0.00258 \\ 
2.70 & 0.35 & 0.06135 & 0.00945 & 0.00614 \\ 
2.70 & 0.45 & 0.07578 & 0.01138 & 0.00758 \\ 
2.70 & 0.55 & 0.09083 & 0.01279 & 0.00908 \\ 
2.70 & 0.65 & 0.13890 & 0.01846 & 0.01389 \\ 
2.70 & 0.75 & 0.17514 & 0.03573 & 0.01751 \\ 
\hline
2.90 & -0.55 & 0.02295 & 0.01465 & 0.00230 \\ 
2.90 & -0.35 & 0.01680 & 0.00546 & 0.00168 \\ 
2.90 & -0.25 & 0.00971 & 0.00376 & 0.00097 \\ 
2.90 & 0.05 & 0.00359 & 0.00222 & 0.00036 \\ 
2.90 & 0.15 & 0.01019 & 0.00284 & 0.00102 \\ 
2.90 & 0.25 & 0.01012 & 0.00403 & 0.00101 \\ 
2.90 & 0.35 & 0.02945 & 0.00543 & 0.00294 \\ 
2.90 & 0.45 & 0.04284 & 0.00702 & 0.00428 \\ 
2.90 & 0.55 & 0.04900 & 0.00873 & 0.00490 \\ 
2.90 & 0.65 & 0.08143 & 0.01379 & 0.00814 \\ 
2.90 & 0.75 & 0.10337 & 0.02153 & 0.01034 \\ \hline
\end{longtable}

\begin{longtable}{|ccccc|}
\caption{G13 differential cross section.}\\
\toprule
{\bf \boldmath$E_{\gamma}$} &
{\bf \boldmath$\cos\theta^{K^{0}}_{CM}$ } &
%{\bf \boldmath$\frac{d\sigma}{d\cos\theta}$} &
{\bf \nicefrac{\boldmath$d\sigma$}{\boldmath$d\cos\theta$}} &
{\bf \boldmath$\delta$(Stat.)} &
{\bf \boldmath$\delta$(Syst.)}\\
{\bf (GeV)} &
{\bf  } &
{\bf (\boldmath$\mu b$)} &
{\bf (\boldmath$\mu b$)} &
{\bf (\boldmath$\mu b$)}\\
\midrule\endfirsthead
\caption[]{(continued) G13 differential cross section.}\\
\toprule
{\bf \boldmath$E_{\gamma}$} &
{\bf \boldmath$\cos\theta^{K^{0}}_{CM}$ } &
%{\bf \boldmath$\frac{d\sigma}{d\cos\theta}$} &
{\bf \nicefrac{\boldmath$d\sigma$}{\boldmath$d\cos\theta$}} &
{\bf \boldmath$\delta$(Stat.)} &
{\bf \boldmath$\delta$(Syst.)}\\
{\bf (GeV)} &
{\bf  } &
{\bf (\boldmath$\mu b$)} &
{\bf (\boldmath$\mu b$)} &
{\bf (\boldmath$\mu b$)}\\ \midrule\endhead
0.97 & -0.65 & 0.75255 & 0.35076 & 0.06867 \\ 
0.97 & -0.55 & 0.77160 & 0.33467 & 0.07040 \\ 
0.97 & -0.45 & 0.86497 & 0.24008 & 0.07892 \\ 
0.97 & -0.35 & 0.74992 & 0.17343 & 0.06843 \\ 
0.97 & -0.25 & 0.72369 & 0.17066 & 0.06603 \\ 
0.97 & -0.15 & 0.59180 & 0.11788 & 0.05400 \\ 
0.97 & -0.05 & 0.70802 & 0.11000 & 0.06460 \\ 
0.97 & 0.05 & 0.78103 & 0.11056 & 0.07126 \\ 
0.97 & 0.15 & 0.64846 & 0.10411 & 0.05917 \\ 
0.97 & 0.25 & 0.57306 & 0.09379 & 0.05229 \\ 
0.97 & 0.35 & 0.57949 & 0.09745 & 0.05287 \\ 
0.97 & 0.45 & 0.70460 & 0.10752 & 0.06429 \\ 
0.97 & 0.55 & 0.44682 & 0.09608 & 0.04077 \\ 
0.97 & 0.65 & 0.51625 & 0.12026 & 0.04710 \\ 
0.97 & 0.75 & 0.51830 & 0.14436 & 0.04729 \\ 
\hline
1.05 & -0.55 & 1.33875 & 0.23334 & 0.10225 \\ 
1.05 & -0.45 & 1.05646 & 0.12457 & 0.08069 \\ 
1.05 & -0.35 & 1.05996 & 0.10028 & 0.08096 \\ 
1.05 & -0.25 & 0.86644 & 0.08602 & 0.06618 \\ 
1.05 & -0.15 & 0.88866 & 0.07516 & 0.06788 \\ 
1.05 & -0.05 & 0.82451 & 0.06602 & 0.06298 \\ 
1.05 & 0.05 & 0.82510 & 0.06255 & 0.06302 \\ 
1.05 & 0.15 & 0.75963 & 0.05883 & 0.05802 \\ 
1.05 & 0.25 & 0.77430 & 0.05987 & 0.05914 \\ 
1.05 & 0.35 & 0.67270 & 0.05781 & 0.05138 \\ 
1.05 & 0.45 & 0.71069 & 0.06013 & 0.05428 \\ 
1.05 & 0.55 & 0.57955 & 0.05932 & 0.04427 \\ 
1.05 & 0.65 & 0.55457 & 0.07076 & 0.04236 \\ 
1.05 & 0.75 & 0.70325 & 0.09656 & 0.05372 \\ 
\hline
1.15 & -0.65 & 1.11443 & 0.18921 & 0.07850 \\ 
1.15 & -0.55 & 1.08909 & 0.12420 & 0.07672 \\ 
1.15 & -0.45 & 1.03075 & 0.08920 & 0.07261 \\ 
1.15 & -0.35 & 0.88525 & 0.06737 & 0.06236 \\ 
1.15 & -0.25 & 0.79475 & 0.05571 & 0.05598 \\ 
1.15 & -0.15 & 0.71598 & 0.04944 & 0.05043 \\ 
1.15 & -0.05 & 0.75928 & 0.04681 & 0.05348 \\ 
1.15 & 0.05 & 0.74950 & 0.04504 & 0.05279 \\ 
1.15 & 0.15 & 0.63681 & 0.03979 & 0.04486 \\ 
1.15 & 0.25 & 0.76139 & 0.04296 & 0.05363 \\ 
1.15 & 0.35 & 0.66951 & 0.04196 & 0.04716 \\ 
1.15 & 0.45 & 0.67394 & 0.04344 & 0.04747 \\ 
1.15 & 0.55 & 0.69452 & 0.04952 & 0.04892 \\ 
1.15 & 0.65 & 0.71138 & 0.05281 & 0.05011 \\ 
1.15 & 0.75 & 0.70978 & 0.06620 & 0.05000 \\ 
\hline
1.25 & -0.65 & 1.41869 & 0.18279 & 0.09557 \\ 
1.25 & -0.55 & 1.02626 & 0.10700 & 0.06913 \\ 
1.25 & -0.45 & 0.92005 & 0.07879 & 0.06198 \\ 
1.25 & -0.35 & 0.80255 & 0.05824 & 0.05406 \\ 
1.25 & -0.25 & 0.76759 & 0.05021 & 0.05171 \\ 
1.25 & -0.15 & 0.70174 & 0.04323 & 0.04727 \\ 
1.25 & -0.05 & 0.72802 & 0.04088 & 0.04904 \\ 
1.25 & 0.05 & 0.62247 & 0.03618 & 0.04193 \\ 
1.25 & 0.15 & 0.65450 & 0.03824 & 0.04409 \\ 
1.25 & 0.25 & 0.77066 & 0.03968 & 0.05191 \\ 
1.25 & 0.35 & 0.77760 & 0.03932 & 0.05238 \\ 
1.25 & 0.45 & 0.72072 & 0.04070 & 0.04855 \\ 
1.25 & 0.55 & 0.62717 & 0.04325 & 0.04225 \\ 
1.25 & 0.65 & 0.67817 & 0.04858 & 0.04568 \\ 
1.25 & 0.75 & 0.68653 & 0.06194 & 0.04625 \\ 
\hline
1.35 & -0.65 & 0.79066 & 0.14608 & 0.05183 \\ 
1.35 & -0.55 & 0.70911 & 0.08701 & 0.04649 \\ 
1.35 & -0.45 & 0.70095 & 0.06063 & 0.04596 \\ 
1.35 & -0.35 & 0.60235 & 0.04678 & 0.03949 \\ 
1.35 & -0.25 & 0.59899 & 0.04013 & 0.03927 \\ 
1.35 & -0.15 & 0.56432 & 0.03514 & 0.03700 \\ 
1.35 & -0.05 & 0.56618 & 0.03263 & 0.03712 \\ 
1.35 & 0.05 & 0.53946 & 0.03094 & 0.03537 \\ 
1.35 & 0.15 & 0.54716 & 0.03028 & 0.03587 \\ 
1.35 & 0.25 & 0.58681 & 0.03221 & 0.03847 \\ 
1.35 & 0.35 & 0.52749 & 0.03141 & 0.03458 \\ 
1.35 & 0.45 & 0.56359 & 0.03440 & 0.03695 \\ 
1.35 & 0.55 & 0.57093 & 0.03933 & 0.03743 \\ 
1.35 & 0.65 & 0.59598 & 0.04497 & 0.03907 \\ 
1.35 & 0.75 & 0.67498 & 0.06028 & 0.04425 \\ 
\hline
1.45 & -0.55 & 0.48669 & 0.07164 & 0.03173 \\ 
1.45 & -0.45 & 0.52218 & 0.05000 & 0.03460 \\ 
1.45 & -0.35 & 0.40599 & 0.03659 & 0.02647 \\ 
1.45 & -0.25 & 0.35432 & 0.02933 & 0.02310 \\ 
1.45 & -0.15 & 0.37073 & 0.02726 & 0.02417 \\ 
1.45 & -0.05 & 0.40146 & 0.02634 & 0.02617 \\ 
1.45 & 0.05 & 0.37243 & 0.02412 & 0.02428 \\ 
1.45 & 0.15 & 0.35336 & 0.02461 & 0.02304 \\ 
1.45 & 0.25 & 0.35558 & 0.02463 & 0.02318 \\ 
1.45 & 0.35 & 0.42516 & 0.02640 & 0.02772 \\ 
1.45 & 0.45 & 0.52429 & 0.03086 & 0.03418 \\ 
1.45 & 0.55 & 0.53086 & 0.03503 & 0.03460 \\ 
1.45 & 0.65 & 0.50279 & 0.03859 & 0.03277 \\ 
1.45 & 0.75 & 0.61126 & 0.05597 & 0.03984 \\ 
\hline
1.55 & -0.55 & 0.28879 & 0.05107 & 0.01964 \\ 
1.55 & -0.45 & 0.24327 & 0.03220 & 0.01633 \\ 
1.55 & -0.35 & 0.20305 & 0.02309 & 0.01362 \\ 
1.55 & -0.25 & 0.19669 & 0.01976 & 0.01320 \\ 
1.55 & -0.15 & 0.18372 & 0.01760 & 0.01233 \\ 
1.55 & -0.05 & 0.17685 & 0.01688 & 0.01193 \\ 
1.55 & 0.05 & 0.23188 & 0.01803 & 0.01559 \\ 
1.55 & 0.15 & 0.27718 & 0.01856 & 0.01865 \\ 
1.55 & 0.25 & 0.30317 & 0.01888 & 0.02034 \\ 
1.55 & 0.35 & 0.31162 & 0.02078 & 0.02092 \\ 
1.55 & 0.45 & 0.32393 & 0.02346 & 0.02173 \\ 
1.55 & 0.55 & 0.38605 & 0.02834 & 0.02589 \\ 
1.55 & 0.65 & 0.43859 & 0.03520 & 0.02941 \\ 
1.55 & 0.75 & 0.43297 & 0.04882 & 0.02904 \\ 
\hline
1.65 & -0.55 & 0.18039 & 0.03551 & 0.01216 \\ 
1.65 & -0.45 & 0.19119 & 0.02561 & 0.01243 \\ 
1.65 & -0.35 & 0.15919 & 0.01849 & 0.01034 \\ 
1.65 & -0.25 & 0.12044 & 0.01560 & 0.00784 \\ 
1.65 & -0.15 & 0.12797 & 0.01427 & 0.00831 \\ 
1.65 & -0.05 & 0.12656 & 0.01391 & 0.00827 \\ 
1.65 & 0.05 & 0.15465 & 0.01498 & 0.01005 \\ 
1.65 & 0.15 & 0.20314 & 0.01597 & 0.01322 \\ 
1.65 & 0.25 & 0.21709 & 0.01689 & 0.01409 \\ 
1.65 & 0.35 & 0.26907 & 0.01870 & 0.01750 \\ 
1.65 & 0.45 & 0.32082 & 0.02180 & 0.02082 \\ 
1.65 & 0.55 & 0.40120 & 0.02820 & 0.02604 \\ 
1.65 & 0.65 & 0.45581 & 0.03610 & 0.02958 \\ 
1.65 & 0.75 & 0.57200 & 0.05412 & 0.03713 \\ 
\hline
1.75 & -0.55 & 0.16534 & 0.03194 & 0.01111 \\ 
1.75 & -0.45 & 0.17897 & 0.02230 & 0.01203 \\ 
1.75 & -0.35 & 0.09317 & 0.01530 & 0.00629 \\ 
1.75 & -0.25 & 0.08305 & 0.01289 & 0.00559 \\ 
1.75 & -0.15 & 0.08291 & 0.01206 & 0.00561 \\ 
1.75 & -0.05 & 0.08655 & 0.01213 & 0.00585 \\ 
1.75 & 0.05 & 0.15390 & 0.01382 & 0.01036 \\ 
1.75 & 0.15 & 0.17531 & 0.01541 & 0.01184 \\ 
1.75 & 0.25 & 0.16149 & 0.01470 & 0.01089 \\ 
1.75 & 0.35 & 0.23698 & 0.01728 & 0.01593 \\ 
1.75 & 0.45 & 0.27890 & 0.02071 & 0.01874 \\ 
1.75 & 0.55 & 0.35337 & 0.02587 & 0.02375 \\ 
1.75 & 0.65 & 0.44786 & 0.03696 & 0.03009 \\ 
1.75 & 0.75 & 0.66910 & 0.06598 & 0.04495 \\ 
\hline
1.85 & -0.55 & 0.06010 & 0.02025 & 0.00409 \\ 
1.85 & -0.45 & 0.08868 & 0.01599 & 0.00592 \\ 
1.85 & -0.35 & 0.09077 & 0.01371 & 0.00609 \\ 
1.85 & -0.25 & 0.08972 & 0.01144 & 0.00600 \\ 
1.85 & -0.15 & 0.08442 & 0.01091 & 0.00564 \\ 
1.85 & -0.05 & 0.08094 & 0.01062 & 0.00550 \\ 
1.85 & 0.05 & 0.09120 & 0.01044 & 0.00611 \\ 
1.85 & 0.15 & 0.11921 & 0.01208 & 0.00797 \\ 
1.85 & 0.25 & 0.13622 & 0.01293 & 0.00913 \\ 
1.85 & 0.35 & 0.18348 & 0.01470 & 0.01226 \\ 
1.85 & 0.45 & 0.27127 & 0.01847 & 0.01812 \\ 
1.85 & 0.55 & 0.27746 & 0.02227 & 0.01853 \\ 
1.85 & 0.65 & 0.36097 & 0.03286 & 0.02411 \\ 
1.85 & 0.75 & 0.45618 & 0.05999 & 0.03046 \\ 
\hline
1.95 & -0.55 & 0.07889 & 0.02144 & 0.00520 \\ 
1.95 & -0.45 & 0.08362 & 0.01455 & 0.00557 \\ 
1.95 & -0.35 & 0.06207 & 0.01126 & 0.00407 \\ 
1.95 & -0.25 & 0.06722 & 0.01060 & 0.00441 \\ 
1.95 & -0.15 & 0.05209 & 0.00908 & 0.00342 \\ 
1.95 & -0.05 & 0.07212 & 0.00919 & 0.00475 \\ 
1.95 & 0.05 & 0.07863 & 0.00974 & 0.00516 \\ 
1.95 & 0.15 & 0.07853 & 0.01053 & 0.00527 \\ 
1.95 & 0.25 & 0.13154 & 0.01223 & 0.00863 \\ 
1.95 & 0.35 & 0.15496 & 0.01353 & 0.01018 \\ 
1.95 & 0.45 & 0.20197 & 0.01687 & 0.01325 \\ 
1.95 & 0.55 & 0.25252 & 0.02099 & 0.01654 \\ 
1.95 & 0.65 & 0.31196 & 0.03198 & 0.02044 \\ 
1.95 & 0.75 & 0.33056 & 0.05756 & 0.02165 \\ 
\hline
2.05 & -0.55 & 0.08179 & 0.01988 & 0.00541 \\ 
2.05 & -0.45 & 0.07376 & 0.01232 & 0.00484 \\ 
2.05 & -0.35 & 0.05625 & 0.01004 & 0.00369 \\ 
2.05 & -0.25 & 0.05190 & 0.00905 & 0.00342 \\ 
2.05 & -0.15 & 0.04399 & 0.00838 & 0.00289 \\ 
2.05 & -0.05 & 0.04490 & 0.00883 & 0.00295 \\ 
2.05 & 0.05 & 0.04106 & 0.00909 & 0.00274 \\ 
2.05 & 0.15 & 0.06679 & 0.00953 & 0.00450 \\ 
2.05 & 0.25 & 0.10552 & 0.01140 & 0.00696 \\ 
2.05 & 0.35 & 0.14621 & 0.01330 & 0.00961 \\ 
2.05 & 0.45 & 0.21218 & 0.01649 & 0.01393 \\ 
2.05 & 0.55 & 0.21077 & 0.02012 & 0.01383 \\ 
2.05 & 0.65 & 0.31704 & 0.03580 & 0.02079 \\ 
2.05 & 0.75 & 0.35042 & 0.06888 & 0.02320 \\ 
\hline
2.15 & -0.55 & 0.05246 & 0.01674 & 0.00358 \\ 
2.15 & -0.45 & 0.04878 & 0.01137 & 0.00327 \\ 
2.15 & -0.35 & 0.03869 & 0.00911 & 0.00263 \\ 
2.15 & -0.25 & 0.04184 & 0.00779 & 0.00281 \\ 
2.15 & -0.15 & 0.02167 & 0.00702 & 0.00147 \\ 
2.15 & -0.05 & 0.02822 & 0.00731 & 0.00190 \\ 
2.15 & 0.05 & 0.04655 & 0.00874 & 0.00313 \\ 
2.15 & 0.15 & 0.04077 & 0.00771 & 0.00276 \\ 
2.15 & 0.25 & 0.07932 & 0.00974 & 0.00546 \\ 
2.15 & 0.35 & 0.11273 & 0.01140 & 0.00756 \\ 
2.15 & 0.45 & 0.14085 & 0.01349 & 0.00944 \\ 
2.15 & 0.55 & 0.24018 & 0.02003 & 0.01609 \\ 
2.15 & 0.65 & 0.24749 & 0.02982 & 0.01658 \\ 
2.15 & 0.75 & 0.25732 & 0.06418 & 0.01761 \\ 
\hline
2.25 & -0.65 & 0.02567 & 0.02226 & 0.00180 \\ 
2.25 & -0.55 & 0.02952 & 0.01342 & 0.00207 \\ 
2.25 & -0.45 & 0.03718 & 0.00992 & 0.00261 \\ 
2.25 & -0.35 & 0.03327 & 0.00812 & 0.00232 \\ 
2.25 & -0.25 & 0.02416 & 0.00730 & 0.00167 \\ 
2.25 & -0.05 & 0.01092 & 0.00615 & 0.00076 \\ 
2.25 & 0.05 & 0.02809 & 0.00760 & 0.00194 \\ 
2.25 & 0.15 & 0.02312 & 0.00776 & 0.00160 \\ 
2.25 & 0.25 & 0.05048 & 0.00915 & 0.00350 \\ 
2.25 & 0.35 & 0.10989 & 0.01185 & 0.00758 \\ 
2.25 & 0.45 & 0.14907 & 0.01409 & 0.01029 \\ 
2.25 & 0.55 & 0.18725 & 0.01990 & 0.01292 \\ 
2.25 & 0.65 & 0.34181 & 0.03665 & 0.02358 \\ 
\hline
2.35 & -0.65 & 0.03675 & 0.02146 & 0.00345 \\ 
2.35 & -0.55 & 0.01943 & 0.01110 & 0.00183 \\ 
2.35 & -0.45 & 0.01774 & 0.00968 & 0.00116 \\ 
2.35 & -0.35 & 0.02601 & 0.00736 & 0.00171 \\ 
2.35 & -0.15 & 0.01446 & 0.00687 & 0.00096 \\ 
2.35 & -0.05 & 0.00745 & 0.00587 & 0.00049 \\ 
2.35 & 0.05 & 0.01752 & 0.00583 & 0.00115 \\ 
2.35 & 0.15 & 0.02863 & 0.00707 & 0.00188 \\ 
2.35 & 0.25 & 0.06421 & 0.00835 & 0.00422 \\ 
2.35 & 0.35 & 0.09907 & 0.01024 & 0.00648 \\ 
2.35 & 0.45 & 0.14870 & 0.01386 & 0.00973 \\ 
2.35 & 0.55 & 0.18374 & 0.01835 & 0.01201 \\ 
2.35 & 0.65 & 0.21345 & 0.03008 & 0.01397 \\ 
\hline
2.45 & -0.55 & 0.02850 & 0.01312 & 0.01366 \\ 
2.45 & -0.45 & 0.02657 & 0.00862 & 0.00176 \\ 
2.45 & -0.35 & 0.01547 & 0.00692 & 0.00104 \\ 
2.45 & -0.25 & 0.01802 & 0.00650 & 0.00118 \\ 
2.45 & -0.15 & 0.01117 & 0.00549 & 0.00090 \\ 
2.45 & -0.05 & 0.00628 & 0.00609 & 0.00042 \\ 
2.45 & 0.05 & 0.00882 & 0.00631 & 0.00057 \\ 
2.45 & 0.15 & 0.02063 & 0.00613 & 0.00135 \\ 
2.45 & 0.25 & 0.04485 & 0.00731 & 0.00292 \\ 
2.45 & 0.35 & 0.06794 & 0.00913 & 0.00442 \\ 
2.45 & 0.45 & 0.11035 & 0.01167 & 0.00718 \\ 
2.45 & 0.55 & 0.16248 & 0.01689 & 0.01057 \\ 
2.45 & 0.65 & 0.19303 & 0.02777 & 0.01260 \\ \hline
\end{longtable}

%\input{g10table.tex}

%\input{g13table.tex}
\begin{comment}
\begin{table*}[tbh]
 %\begin{minipage}{\textwidth}
\centering
\caption{Total Cross Section from g13 dataset}
\label{tab:totcross1}
\begin{tabular}{|c|c|c|c|c|}
\hline
 {\bf Energy} &  {\bf $d\sigma(E_{\gamma})$ ($\mu b$)} & {\bf Statistical} & {\bf Systematic)} \\ 
  {\bf } &  {\bf} & {\bf Uncert($\mu b$)} & {\bf Uncert($\mu b$)} \\ \hline
 0.97 &  1.273  &  0.188  &  0.029   \\ \hline
 1.05 &  1.776  &  0.155  &  0.273   \\ \hline
 1.15 &  1.755  &  0.083  &  0.068   \\ \hline
 1.25 &  1.786  &  0.074  &  0.189   \\ \hline
 1.35 &  1.281  &  0.060  &  0.006   \\ \hline
 1.45 &  1.036  &  0.058  &  0.058   \\ \hline
 1.55 &  0.686  &  0.041  &  0.080   \\ \hline
 1.65 &  0.619  &  0.032  &  0.034   \\ \hline
 1.75 &  0.559  &  0.029  &  0.018   \\ \hline
 1.85 &  0.385  &  0.021  &  0.114   \\ \hline
 1.95 &  0.349  &  0.021  &  0.039   \\ \hline
 2.05 &  0.350  &  0.019  &  0.050   \\ \hline
 2.15 &  0.275  &  0.017  &  0.074   \\ \hline
 2.25 &  0.268  &  0.015  &  0.047   \\ \hline
 2.35 &  0.230  &  0.014  &  0.036   \\ \hline
 2.45 &  0.198  &  0.014  &  0.022   \\ \hline
\end{tabular}
%\end{minipage}
\end{table*}
\end{comment}
\clearpage
%\FloatBarrier
%_____________________________REFERENCES________________________
\bibliography{k0LambdaPub}

\end{document}